\documentclass[twocolumn]{aastex631}

\usepackage{amsmath}
\usepackage{empheq}
\usepackage{mathrsfs}
\usepackage{textcomp}
\usepackage{enumitem}   
\usepackage{gensymb}
\usepackage{hyperref}
\usepackage{graphicx}
\usepackage[caption=false]{subfig}
\usepackage{multirow}
\usepackage{longtable}
\usepackage[flushleft]{threeparttable}
\usepackage{booktabs} 
\usepackage{CJK}
\hypersetup{colorlinks, linkcolor={blue}, citecolor={blue}, urlcolor={blue}} 

\definecolor{DarkOrange}{RGB}{204, 85, 0}
\definecolor{LincolnGreen}{RGB}{17, 102, 0}
\def\ion#1#2{#1$\;${\footnotesize\rm{#2}}\relax}

\newcommand{\ad}[1]{{\color{black} {#1}}}

\usepackage{xspace}

\newcommand\nustar{\textit{NuSTAR}\xspace}
\newcommand\nicer{\textit{NICER}\xspace}
\newcommand\chandra{\textit{Chandra}\xspace}
\newcommand\swift{\textit{Swift}\xspace}
\newcommand\xmm{\textit{XMM-Newton}\xspace}

\newcommand\srg{\textit{SRG}\xspace}
\newcommand\srge{\textit{SRG}/eROSITA\xspace}
\newcommand\rosat{\textit{ROSAT}\xspace}

\newcommand\cps{$\rm count\,s^{-1}$\xspace}

\newcommand\target{AT2021ehb\xspace}

\def \caltech {{Division of Physics, Mathematics and Astronomy, 
California Institute of Technology, Pasadena, CA 91125, USA}}

\def \coo {{Caltech Optical Observatories, California Institute of Technology, Pasadena, CA 91125, USA}}

\def \gsfc {{Astrophysics Science Division, NASA Goddard Space Flight Center, Greenbelt, MD 20771, USA}}

\def \iki {{Space Research Institute, Russian Academy of Sciences, Profsoyuznaya ul. 84/32, Moscow, 117997, Russia}}

\def \mpa {{Max-Planck-Institut f\"{u}r Astrophysik, Karl-Schwarzschild-Str. 1, D-85741 Garching, Germany}}

\def\msun{M_\odot}
\def\mr{\mathrm}

\begin{document}
\pagenumbering{arabic}

\title{The Tidal Disruption Event AT2021ehb: Evidence of Relativistic Disk Reflection,\\
and Rapid Evolution of the Disk--Corona System
}

\correspondingauthor{Yuhan Yao}
\email{yyao@astro.caltech.edu}

\author[0000-0001-6747-8509]{Yuhan Yao}
\affiliation{\caltech}

\author[0000-0002-1568-7461]{Wenbin Lu}
\affiliation{Department of Astrophysical Sciences, 4 Ivy Lane, Princeton University, Princeton, NJ 08544, USA}
\affiliation{Department of Astronomy, University of California, Berkeley, CA 94720-3411, USA}

\author[0000-0002-5063-0751]{Muryel Guolo}
\affiliation{Department of Physics and Astronomy, Johns Hopkins University, 3400 N. Charles Street, Baltimore, MD 21218, USA}

\author[0000-0003-1386-7861]{Dheeraj R. Pasham}
\affiliation{Kavli Institute for Astrophysics and Space Research, Massachusetts Institute of Technology, Cambridge, MA 02139, USA}

\author[0000-0003-3703-5154]{Suvi Gezari}
\affiliation{Space Telescope Science Institute, 3700 San Martin Drive, Baltimore, MD 21218, USA}

\author{Marat Gilfanov}
\affiliation{\iki}
\affiliation{\mpa}

\author{Keith C. Gendreau}
\affiliation{\gsfc}

\author[0000-0003-2992-8024]{Fiona Harrison}
\affiliation{\caltech}

\author[0000-0003-1673-970X]{S. Bradley Cenko}
\affiliation{\gsfc}

\author[0000-0001-5390-8563]{S. R. Kulkarni}
\affiliation{\caltech}

\author[0000-0003-2869-7682]{Jon M. Miller}
\affiliation{Department of Astronomy, The University of Michigan, 1085 South University Avenue, Ann Arbor, Michigan 48103, USA}

\author[0000-0001-5819-3552]{Dominic J. Walton}
\affiliation{Centre for Astrophysics Research, University of Hertfordshire, College Lane, Hatfield AL10 9AB, UK}
\affiliation{Institute of Astronomy, University of Cambridge, Madingley Road, Cambridge CB3 0HA, UK}

\author[0000-0003-3828-2448]{Javier~A.~Garc\'ia}
\affiliation{\caltech}
\affiliation{Dr. Karl Remeis-Observatory and Erlangen Centre for Astroparticle Physics, Sternwartstr.~7, 96049 Bamberg, Germany}

\author[0000-0002-3859-8074]{Sjoert van Velzen}
\affiliation{Leiden Observatory, Leiden University, Postbus 9513, 2300 RA, Leiden, The Netherlands}

\author[0000-0002-8297-2473]{Kate D. Alexander}
\affiliation{Center for Interdisciplinary Exploration and Research in Astrophysics (CIERA) and Department of Physics and Astronomy, Northwestern University, Evanston, IL 60208, USA}

\author[0000-0003-3124-2814]{James C. A. Miller-Jones}
\affiliation{International Centre for Radio Astronomy Research Curtin University, GPO Box U1987, Perth, WA 6845, Australia}

\author[0000-0002-2555-3192]{Matt Nicholl}
\affiliation{Birmingham Institute for Gravitational Wave Astronomy and School of Physics and Astronomy, University of Birmingham, Birmingham B15 2TT, UK}

\author[0000-0002-5698-8703]{Erica Hammerstein}
\affiliation{Department of Astronomy, University of Maryland, College Park, MD 20742, USA}

\author[0000-0002-9380-8708]{Pavel Medvedev}
\affiliation{\iki}

\author[0000-0003-2686-9241]{Daniel Stern}
\affiliation{Jet Propulsion Laboratory, California Institute of Technology, 4800 Oak Grove Drive, Pasadena, CA 91109, USA}

\author[0000-0002-7252-5485]{Vikram Ravi}
\affiliation{\caltech}

\author{R. Sunyaev}
\affiliation{\iki}
\affiliation{\mpa}


\author[0000-0002-7777-216X]{Joshua S. Bloom}
\affiliation{Department of Astronomy, University of California, Berkeley, CA 94720-3411, USA}
\affiliation{Lawrence Berkeley National Laboratory, 1 Cyclotron Road, MS 50B-4206, Berkeley, CA 94720, USA}

\author[0000-0002-3168-0139]{Matthew J. Graham}
\affiliation{\caltech}

\author[0000-0002-7252-3877]{Erik C. Kool}
\affiliation{The Oskar Klein Centre, Department of Astronomy, Stockholm University, AlbaNova, SE-10691, Stockholm, Sweden}

\author[0000-0003-2242-0244]{Ashish~A.~Mahabal}
\affiliation{\caltech}
\affiliation{Center for Data Driven Discovery, California Institute of Technology, Pasadena, CA 91125, USA}

\author[0000-0002-8532-9395]{Frank J. Masci}
\affiliation{IPAC, California Institute of Technology, 1200 E. California Blvd, Pasadena, CA 91125, USA}

\author[0000-0003-1227-3738]{Josiah Purdum}
\affiliation{\coo}

\author[0000-0001-7648-4142]{Ben Rusholme}
\affiliation{IPAC, California Institute of Technology, 1200 E. California Blvd, Pasadena, CA 91125, USA}

\author[0000-0003-4531-1745]{Yashvi Sharma}
\affiliation{\caltech}

\author[0000-0001-7062-9726]{Roger Smith}
\affiliation{\coo}

\author[0000-0003-1546-6615]{Jesper Sollerman}
\affiliation{The Oskar Klein Centre, Department of Astronomy, Stockholm University, AlbaNova, SE-10691, Stockholm, Sweden}

\begin{abstract}
We present X-ray, UV, optical, and radio observations of the nearby ($\approx$78\,Mpc) tidal disruption event (TDE) \target/ZTF21aanxhjv during its first 430\,days of evolution. 
\target occurs in the nucleus of a galaxy hosting a $\approx 10^{7}\,M_\odot$ black hole ($M_{\rm BH}$ inferred from host galaxy scaling relations). High-cadence \swift and \nicer monitoring reveals a delayed X-ray brightening. The spectrum first undergoes a gradual ${\rm soft }\rightarrow{\rm hard}$ transition and then suddenly turns soft again within 3 days at $\delta t\approx 272\,$days during which the X-ray flux drops by a factor of ten.
In the joint \textit{NICER}+\nustar observation ($\delta t =264$\,days, harder state), we observe a prominent non-thermal component up to 30\,keV and an extremely broad emission line in the iron $K$ band.
The bolometric luminosity of \target reaches a maximum of $6.0^{+10.4}_{-3.8}\% L_{\rm Edd}$ when the X-ray spectrum is the hardest. 
During the dramatic X-ray evolution, no radio emission is detected, the UV/optical luminosity stays relatively constant, and the optical spectra are featureless.
We propose the following interpretations: (i) the ${\rm soft }\rightarrow{\rm hard}$ transition may be caused by the gradual formation of a magnetically dominated corona; (ii) hard X-ray photons escape from the system along solid angles with low scattering optical depth ($\sim\,$a few) whereas the UV/optical emission is likely generated by reprocessing materials with much larger column density --- the system is highly aspherical; (iii) the abrupt X-ray flux drop may be triggered by the thermal-viscous instability in the inner accretion flow, leading to a much thinner disk.
\end{abstract}
\keywords{
Tidal disruption (1696);
X-ray transient sources (1852); 
Supermassive black holes (1663);
Time domain astronomy (2109); 
High energy astrophysics (739); 
Accretion (14)
}

\vspace{1em}

\section{Introduction}
A star getting too close to a massive black hole (MBH) can get disrupted by the tidal forces in a Tidal Disruption Event (TDE; see recent review by \citealt{Gezari2021}). 
The first observational evidence for TDEs came from the detection of X-ray flares from the centers of quiescent galaxies during the \rosat (0.1--2.4\,keV) all-sky survey (RASS) in 1990--1991 \citep{Donley2002}. The flares exhibit soft spectra that are consistent with blackbody radiation with temperatures $T_{\rm bb}\sim 10^6\,{\rm K}$ and radii $R_{\rm bb}\sim {\rm few}\times 10^{11}\,{\rm cm}$ \citep{Saxton2020}. Since 2020, the \textit{Spektrum-Roentgen-Gamma} (\srg) mission \citep{Sunyaev2021}, with its sensitive  eROSITA telescope (0.2--8\,keV; \citealt{Predehl2021}) and six month cadenced all-sky surveys, has become the most prolific discoverer of TDEs in X-rays. The majority of X-ray selected TDEs are faint in the optical \citep{Sazonov2021}.

In the UV and optical sky, TDEs have been identified as blue nuclear transients in surveys such as the \textit{Galaxy Evolution Explorer} \citep{Martin2005}, the Panoramic Survey Telescope and Rapid Response System DR1 (Pan-STARRS, PS1; \citealt{Flewelling2020, Waters2020}), the Sloan Digital Sky Survey (SDSS, \citealt{Alam2015}), the All-Sky Automated Survey for SuperNovae (ASAS-SN; \citealt{Shappee2014}), the Palomar Transient Factory (PTF; \citealt{Law2009, Rau2009}), the intermediate PTF (iPTF), the Asteroid Terrestrial-impact Last Alert System (ATLAS; \citealt{Tonry2018}), and the Zwicky Transient Facility (ZTF; \citealt{Bellm2019b, Graham2019}). 
In most cases, the UV/optical spectral energy distribution (SED) can be described by blackbody radiation with larger radii ($R_{\rm bb}\sim {\rm few}\times 10^{14}\,{\rm cm}$) and lower temperatures ($T_{\rm bb}\sim{\rm few}\times 10^4\,{\rm K}$) than those of the X-ray discovered events.
The origin of this blackbody component has been attributed to reprocessing of disk emission by an optically thick gas layer \citep{Metzger2016, Roth2016, Lu2020}, stream self-intersecting shocks formed as a result of general relativistic apsidal precession \citep{Piran2015, Jiang2016_self_crossing_shock}, or intrinsic thermal emission from the viscously heated accretion disk \citep{Wevers2021}. 

Among the UV/optically selected TDEs with simultaneous X-ray observations, about two dozen events have been detected in the X-rays \ad{(e.g., \citealt{Auchettl2017, Wevers2020})}. Their X-ray light curves show a wide range of properties. For example, the X-ray emission of ASASSN-14li lags behind its UV/optical emission by one month \citep{Pasham2017}; ASASSN-15oi, AT2018fyk, and AT2019azh exhibit a gradual X-ray brightening long after the UV/optical peak \citep{Gezari2017, Wevers2021, Hinkle2021}; AT2019ehz and OGLE16aaa show extreme X-ray flares on a timescale of a few days \citep{vanVelzen2021, Kajava2020, Shu2020}; and the probable neutrino emitter AT2019dsg has a rapid X-ray decline \citep{Stein2021}. Understanding the co-evolution between the X-ray and UV/optical emission may hold the key in deciphering the origin of these two components.

The majority of TDEs are not associated with on-axis relativistic jets \citep{Alexander2020} (hereafter non-jetted TDEs). 
The sample of jetted TDEs includes four objects: Sw\,J1644+57 \citep{Bloom2011, Burrows2011, Zauderer2011}, Sw\,J2058+05 \citep{Cenko2012, Pasham2015}, and Sw\,J1112$-$82 \citep{Brown2015} were discovered by the hard X-ray Burst Alert Telescope (BAT) on board \swift, whereas AT2022cmc was discovered by ZTF in the optical \citep{Andreoni2022_gcn_discovery, Yao2022_ATel_22cmc, Pasham2022_ATel_22cmc}. Among them, Sw\,J1644+57 is the most well studied. Its fast X-ray variability and extremely high isotropic equivalent X-ray luminosity ($\sim 10^{47}\,{\rm erg\,s^{-1}}$) suggest that the early-time X-rays are powered by internal dissipation within a jet. A sudden X-ray flux drop by a factor of $\sim10^2$ indicates a jet shut off at rest-frame 370\,days after discovery \citep{Zauderer2013}, after which the X-ray emission is consistent with being powered by a forward shock \citep{Eftekhari2018, Cendes2021a}. 

During the outburst of a stellar-mass black hole X-ray binary (XRB), as the mass accretion rate ($\dot M_{\rm acc}$) varies, the X-ray source transitions between distinct spectral states governed by the global evolution of the disk--corona system \citep{Remillard2006}. A major question in accretion physics is whether a similar geometry operates in the environment around MBHs. Recent studies of a sample of Changing-Look Active Galactic Nuclei (CLAGNs) support a scale-invariant nature of black hole accretion flows \citep{McHardy2006, Walton2012, Ruan2019}. However, the preexisting gas and dusty torus sometimes complicate interpretation of the observables in CLAGNs \citep{Guolo2021}. On the other hand, the majority of TDEs are hosted by otherwise quiescent galaxies \citep{French2020}. Therefore, TDEs provide ideal laboratories for studying MBH accretion in different regimes \citep{Ulmer1999, Strubbe2009}. 

ZTF conducts multiple time-domain surveys using the ZTF mosaic camera \citep{Dekany2020} on the Palomar Oschin Schmidt 48-inch (P48) telescope. The ZTF team selects TDE candidates by imposing a set of criteria, such as proximity to a galaxy nucleus, a lack of pre-flare nuclear activity, a lack of $g-r$ color change, etc (see detailed descriptions in \citealt{vanVelzen2019, vanVelzen2021}). 
The filter is executed by AMPEL \citep{Nordin2019}. We use the Fritz marshal\footnote{\url{https://github.com/fritz-marshal/fritz}} to coordinate our follow-up classifications.
Thanks to its fast survey speed, ZTF is now reporting $\sim15$ TDEs per year \citep{vanVelzen2021, Hammerstein2022}. 

\target/ZTF21aanxhjv was first detected by the ZTF public 2-day cadence all-sky survey at a brightness of $g_{\rm ZTF} = 19.10\pm0.22$ on 2021 March 1. 
On 2021 March 3, it was reported to the Transient Name Server (TNS) by the ALeRCE broker \citep{Munoz-Arancibia2021}.
On 2021 March 25, \target passed our TDE selection filter.
\swift observations were triggered while the TDE was still on the rise to peak. 
On 2021 March 26, we classified \target as a TDE based on its nuclear location, persistent blue color, and bright UV emission \citep{Gezari2021_ehb_discovery, Yao2021_CR_21ehb}. 
Four \swift snapshots from 2021 March 26 to April 2 yielded no X-ray detections.
From 2021 April 12 to June 16, \target was not observed due to occultation by the Sun. On 2021 June 17, ZTF observations resumed. On 2021 July 1, X-rays were detected with \swift \citep{Yao2021_ehbXray}. 
Its bright X-ray emission ($\sim10^{42}\,{\rm erg\,s^{-1}}$) and the subsequent X-ray brightening motivated us to conduct a comprehensive monitoring campaign.

At a spectroscopic redshift of $z=0.0180$ (see \S\ref{subsec:esi}), \target is the third closest TDE discovered by optical sky surveys. 
The previously known lower-redshift events, AT2019qiz \citep{Nicholl2020} and iPTF16fnl \citep{Blagorodnova2017}, were too faint in the X-ray to be carefully characterized.
\target, with a peak 0.3--10\,keV X-ray flux of 1\,mCrab, is the brightest non-jetted TDE in the X-ray sky. We are therefore able to conduct high-cadence monitoring (with \swift and \nicer) and obtain high signal-to-noise ratio (SNR) X-ray spectra (with \nustar, \nicer, \xmm, and \srge), which allows for the search of spectral line features in the X-ray continuum.

Unlike the X-ray spectra of most other non-jetted TDEs \citep{Saxton2020, Sazonov2021}, the X-ray spectrum of \target drastically evolves over the X-ray observing campaign of $\sim 370$\,days, and at a certain stage exhibits prominent non-thermal hard emission.
Therefore, \target is only the second non-jetted TDE, after AT2018fyk \citep{Wevers2021}, which allows us to investigate the rapid evolution between the UV/optical, soft X-ray, and hard X-ray components. Different from the result presented by \citet{Wevers2021}, we find that the disk--corona system of \target is dissimilar to XRBs.

In this paper, we present an in-depth study of the X-ray, UV, optical, and radio emission of AT2021ehb, using observations obtained from 2021 March 1 to 2022 May 31. 
We outline the observations in \S\ref{sec:obs}. 
We analyze the host galaxy in \S\ref{sec:host}, including measurements of the central black hole mass ($M_{\rm BH}$) and the SED.
We study the light curve and spectral evolution of the TDE emission in \S\ref{sec:TDEanalysis}.
We provide a discussion in \S\ref{sec:discussion}, and conclude in \S\ref{sec:conclusion}.

UT time is used throughout the paper. We adopt a standard $\Lambda$CDM cosmology with matter density $\Omega_{\rm M} = 0.3$, dark energy density $\Omega_{\Lambda}=0.7$, and the Hubble constant $H_0=70\,{\rm km\,s^{-1}\,Mpc^{-1}}$, implying a luminosity distance to \target of $D_L = 78.2\,{\rm Mpc}$. 
UV and optical magnitudes are reported in the AB system. 
We use the extinction law from \citet{Cardelli1989}, and adopt a Galactic extinction of $E_{B-V, {\rm MW}} =  0.123$\,mag \citep{Schlafly2011}. 
Uncertainties of X-ray model parameters are reported at the 90\% confidence level. Other uncertainties are 68\% confidence intervals, and upper limits are reported at $3\sigma$. 
Coordinates are given in J2000. 

\section{Observations and Data Reduction} \label{sec:obs}

\subsection{ZTF Optical Photometry}\label{subsec:ztf}
We obtained ZTF\footnote{\url{https://ztfweb.ipac.caltech.edu/cgi-bin/requestForcedPhotometry.cgi}} forced photometry \citep{Masci2019} in the $g$ and the $r$ bands using the median position of all ZTF alerts up to MJD 59550 ($\alpha=03$h07m47.82s, $\delta=+40^{\circ}18^{\prime}40.85^{\prime\prime}$). We performed baseline correction following the procedures outlined in \citet{Yao2019}. 

The peak of the optical light curve probably occurred during Sun occultation and cannot be robustly determined. Therefore, we fitted a five-order polynomial function to the $r_{\rm ZTF}$-band observations, which suggested that the optical maximum light was around $ {\rm MJD} \approx 59321$. Hereafter we use $\delta t$ to denote rest-frame days relative to MJD 59321. 
The Galactic extinction-corrected ZTF light curves are shown in Figure~\ref{fig:lc_uvopt}. All ZTF photometry is provided in Appendix~\ref{subsec:obs_log} (Table~\ref{tab:phot}).

\begin{figure}[htbp!]
    \centering
    \includegraphics[width = \columnwidth]{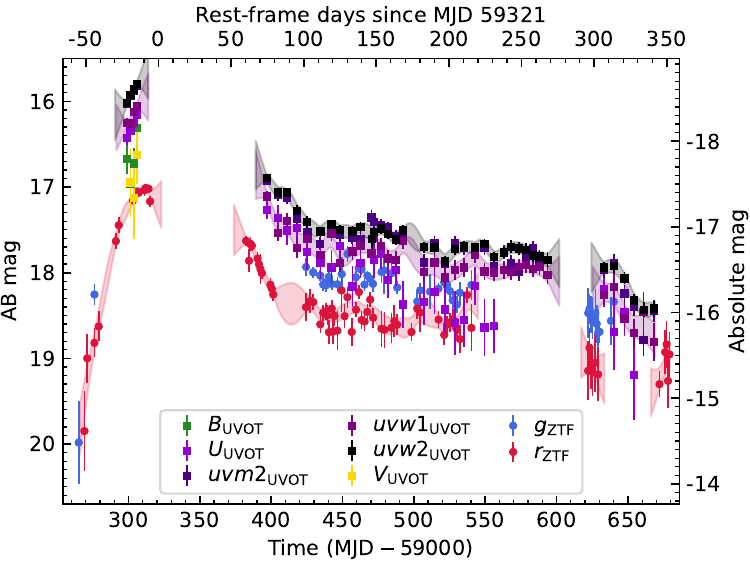}
    \caption{Optical and UV light curves of \target. 
    The host contribution has been removed using difference photometry (ZTF, \S\ref{subsec:ztf}) or subtraction of fluxes estimated from the galaxy SED (UVOT, \S\ref{subsubsec:uvot}). 
    Photometry has only been corrected for Galactic extinction.
    The transparent lines are simple Gaussian process fits in each filter (see \S\ref{subsec:uvopt_bbfit}), where the width of the lines represent $1\sigma$ model uncertainties. 
    For clarity, we only show the model fits in the $r_{\rm ZTF}$, $uvw1$, and $uvw2$ bands. 
    Regions where the model uncertainty is greater than 0.3\,mag are not shown.
    The lack of ZTF and UVOT data at $0 \lesssim \delta t \lesssim 50$\,days is due to Sun occultation; 
    The lack ZTF data at $220 \lesssim \delta t \lesssim 290$\,days and $310 \lesssim \delta t \lesssim 340$\,days is due to performance issues with the cooling system for the ZTF Camera \citep{Fremling2021}; 
    The lack of UVOT data at $270 \lesssim \delta t \lesssim 300$\,days is due to an issue with one of the \swift reaction wheels \citep{Cenko2022}. 
    \label{fig:lc_uvopt}}
\end{figure}

\subsection{SEDM and LT Optical Photometry}
We obtained additional $ugri$ photometry using the Spectral Energy Distribution Machine (SEDM, \citealt{Blagorodnova2018}, \citealt{Rigault2019}) on the robotic Palomar 60 inch telescope (P60, \citealt{Cenko2006}), and the optical imager (IO:O) on the Liverpool Telescope (LT; \citealt{Steele2004}). The SEDM photometry was host-subtracted using the automated pipeline \texttt{FPipe} \citep{Fremling2016}. The LT photometry was host subtracted using SDSS images.

We found a mismatch between the SEDM/LT $gr$ photometry and the ZTF photometry. 
This is probably a result of different reference images being used. 
The ZTF difference photometry is more reliable since the reference images were constructed using P48 observations taken in 2018--2019. The reference images of SEDM/LT come from SDSS images (taken in 2005), and long-term variability of the galaxy nucleus will render the difference photometry less robust. Therefore, we present the SEDM and LT photometry in Appendix~\ref{subsec:obs_log} (Table~\ref{tab:phot}), but exclude them in the following analysis.

\subsection{Optical Spectroscopy}

\begin{figure*}[htbp!]
    \centering
    \includegraphics[width=0.95\textwidth]{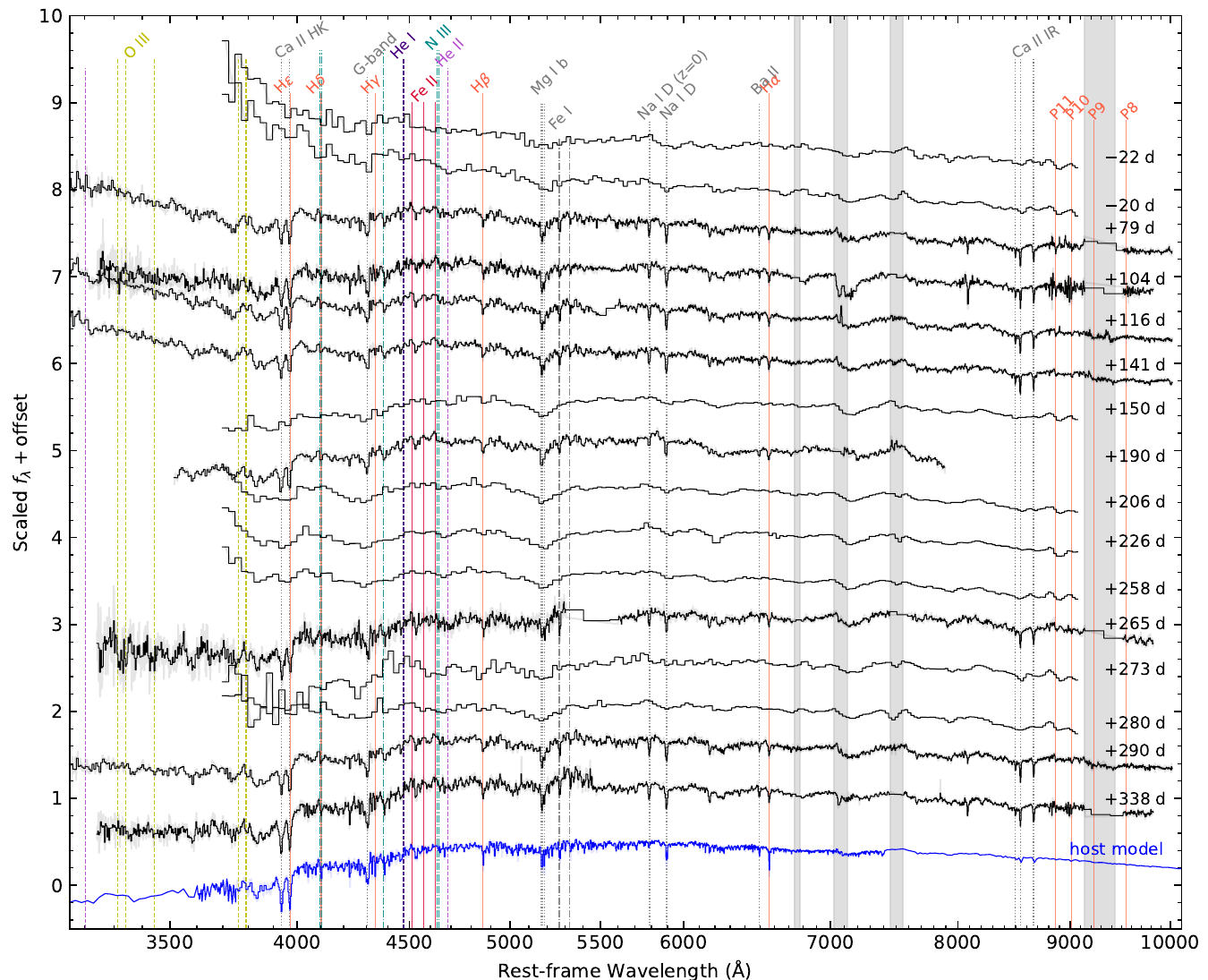}
    \caption{Optical spectroscopic evolution of \target. The observed spectra have been corrected for Galactic extinction. The vertical lines mark observed strong host absorption lines and spectral features common in TDEs. 
    The vertical grey bands mark atmospheric telluric features and strong telluric features have been masked. 
    The best-fit galaxy model is shown at the bottom (see \S\ref{subsec:host_fsps}).
    \label{fig:opt_spec}}
\end{figure*}

We obtained low-resolution optical spectroscopic observations using
the Low Resolution Imaging Spectrograph (LRIS; \citealt{Oke1995}) on the Keck-I telescope,
the Double Spectrograph (DBSP; \citealt{Oke1982}) on the 200-inch Hale telescope, 
the integral field unit (IFU; $R\approx 100$) spectrograph of SEDM,
and the De Veny Spectrograph on the Lowell Discovery Telescope (LDT). 
We also obtained a medium-resolution spectrum using the Echellette Spectrograph and Imager (ESI; \citealt{Sheinis2002}) on the Keck-II telescope.

Figure~\ref{fig:opt_spec} shows the low-resolution spectra. 
The instrumental details and an observing log can be found in Appendix~\ref{sec:optspec_details}.

\subsection{\swift}
\target was observed by the X-Ray Telescope (XRT; \citealt{Burrows2005}) and the Ultra-Violet/Optical Telescope (UVOT; \citealt{Roming2005}) on board \swift under our GO program 1619088 (as ZTF21aanxhjv; target ID 14217; PI: Gezari) and a series of time-of-opportunity (ToO) requests (PI: Yao). All \swift data were processed with \texttt{heasoft} v6.29c.

\subsubsection{XRT} \label{subsubsec:XRTobs}
All XRT observations were obtained in the photon-counting mode. 
First, we ran \texttt{ximage} to select snapshots where \target was detected above $3\sigma$. 
For X-ray non-detections, we computed upper limits within a circular region with a radius of $30^{\prime\prime}$, assuming Poisson statistics. 
For X-ray detections, to calculate the background-subtracted count rates, we filtered the cleaned event files using a source region with $r_{\rm src} = 30^{\prime\prime}$, and eight background regions with $r_{\rm bkg} = 25^{\prime\prime}$ evenly spaced at $80^{\prime\prime}$ from \target. A log of XRT observations is given in Appendix~\ref{subsec:obs_log} (Table~\ref{tab:xrt}).

We generated XRT spectra using an automated online tool\footnote{\url{https://www.swift.ac.uk/user_objects}} \citep{Evans2009}. To improve the SNR of each spectrum, we stacked consecutive observations with a similar hardness ratio (HR; see details in \S\ref{subsubsec:XRTspec}). 

\subsubsection{UVOT} \label{subsubsec:uvot}
The first four UVOT epochs (obsID 14217001--14217005) were conducted with $UBV$+All UV filters. Subsequent observations were 
conducted with $U$+All UV filters. 

We measured the UVOT photometry using the \texttt{uvotsource} tool. We used a circular source region with $r_{\rm src} = 12^{\prime\prime}$, and corrected for the enclosed energy within the aperture\footnote{A large aperture is chosen to make sure that all of the flux of the host galaxy is captured.}. We measured the background using two nearby circular source-free regions with $r_{\rm bkg} = 15^{\prime\prime}$.
Following the procedures outlined in \citet{vanVelzen2021}, we estimated the host-galaxy flux in the UVOT bandpass from the population synthesis models (see \S\ref{subsec:host_fsps}). The UVOT light curves are presented in Figure~\ref{fig:lc_uvopt} and provided in Appendix~\ref{subsec:obs_log} (Table~\ref{tab:phot}).

\subsection{\nicer} \label{subsec:NICERobs}
\target was observed by the Neutron Star Interior Composition Explorer (\nicer; \citealt{Gendreau2016}) under Director's Discretionary Time (DDT) programs on 2021 March 26, 2021 July 2--7, and from 2021 November 13 to 2022 March 29 (PIs: Yao, Gendreau, Pasham). 
The \nicer data were processed using \texttt{nicerdas} v9 (\texttt{2021-08-31\_V008c}). 
We ran \texttt{nicerl2} to obtain the cleaned and screened event files. Background was computed using the \texttt{nibackgen3C50} tool \citep{Remillard2022}. Following the screening criteria suggested by \citet{Remillard2022}, we removed GTIs with \texttt{hbgcut=0.05} and \texttt{s0cut=2.0}.

We extracted one spectrum for each obsID, excluding obsIDs with 0.3--1\,keV background rate $>0.2$\,\cps or 4--12\,keV background rate $>0.1$\,\cps. 
Using observations bracketed by the two \nustar observations, we also produced two \nicer spectra with exposure times of 8.2\,ks and 36.6\,ks, which we jointly analyzed with the \nustar spectra (see \S\ref{subsubsec:joint1} and \S\ref{subsubsec:joint2}).

All \nicer spectra were binned using the optimal binning scheme \citep{Kaastra2016}, requiring at least 20 counts per bin.
Following the \nicer calibration memo\footnote{See \url{https://heasarc.gsfc.nasa.gov/docs/nicer/data_analysis/nicer_analysis_tips.html}.}, we added systematic errors of 1.5\% with \texttt{grppha}.

\subsection{\xmm}

We obtained two epochs of follow-up observations with \xmm under our Announcement of Opportunity (AO) program (PI: Gezari), on 2021 August 4 (obsID 0882590101) and 2022 January 25 (obsID 0882590901). The observations were taken in Full Frame mode with the thin filter using the European Photon Imaging Camera (EPIC; \citealt{Struder2001}).

The observation data files (ODFs) were reduced using the \xmm Standard Analysis Software \citep{Gabriel_04}.
The raw data files were then processed using the \texttt{epproc} task. 
Since the pn instrument generally has better sensitivity than MOS1 and MOS2, we only analyze the pn data. 
Following the \xmm data analysis guide, to check for background activity and generate ``good time intervals'' (GTIs), we manually inspected the background light curves in the 10--12\,keV band. 
Using the \texttt{evselect} task, we only retained patterns that correspond to single and double events (\texttt{PATTERN<=4}). 

The source spectra were extracted using a source region of $r_{\rm src} = 35^{\prime\prime}$ around the peak of the emission. 
The background spectra were extracted from a $r_{\rm bkg} = 108^{\prime\prime}$ region located in the same CCD.
The ARFs and RMF files were created using the \texttt{arfgen} and \texttt{rmfgen} tasks, respectively. We grouped the spectra to have at least 25 counts per bin, and limited the over-sampling of the instrumental resolution to a factor of 5.

\subsection{\srge}

\begin{deluxetable}{ccrc}[htbp!]
    \tablecaption{Log of \srg observations of \target. \label{tab:srg}}
	\tablehead{
	\colhead{eRASS}
	& \colhead{MJD}
	& \colhead{$\delta t$}
	& \colhead{0.3--10\,keV flux} \\
	\colhead{}
	& \colhead{}
	& \colhead{(days)}
	& \colhead{($10^{-13}\,{\rm erg\,s^{-1}\,cm^{-2}}$)}
	}
	\startdata
    1 & 58903.59--58904.59 & $-409.5$ & $<0.25$ \\
    2 & 59083.36--59084.70 & $-232.8$ & $<0.23$  \\
    3 & 59253.16--59254.16 & $-66.1$ & $<0.23$ \\
    4 & 59442.45--59443.62 & $+119.9$  & $76.8^{+2.5}_{-2.4}$ \\
    5 & 59624.53--59625.70 & $+298.7$ & $30.7^{+2.4}_{-2.3}$\\
	\enddata
	\tablecomments{Upper limits are computed assuming an absorbed power-law spectrum with $\Gamma=2.5$ and $N_{\rm H} = 9.97\times 10^{20}\,{\rm cm^{-2}}$, and presented at 90\% confidence. }
\end{deluxetable}

The location of AT2021ehb was scanned by eROSITA 
as part of the planned eight all-sky surveys. Hereafter eRASS$n$ refers to the $n$'th eROSITA all-sky survey\footnote{Here $n$ runs from 1 to 8. As of April 2022, eRASS1--eRASS4 have been completed, and 38\% (sky area) of eRASS5 has been completed.}. During eRASS4, AT2021ehb was independently identified by \srg as a TDE candidate. A log of \srg observations is given in Table~\ref{tab:srg}. We grouped the eRASS4 and eRASS5 spectra to have at least 3 counts per bin.

\subsection{\nustar}
We obtained Nuclear Spectroscopic Telescope ARray (\nustar; \citealt{Harrison2013}) observations under a pre-approved ToO program (PI: Yao; obsID 80701509002) and a DDT program (PI: Yao; obsID 90801501002). 
The first epoch was conducted from 2021 November 18.8 to 19.9 with an exposure time of 43.2\,ks.
The second epoch was conducted from 2022 January 10.4 to 12.1 with an exposure time of 77.5\,ks.

To generate the first epoch's spectra for the two photon-counting detector modules (FPMA and FPMB), source photons were extracted from a circular region with a radius of $r_{\rm src}=40^{\prime\prime}$ centered on the apparent position of the source in both FPMA and FPMB. The background was extracted from a $r_{\rm bkg}=80^{\prime\prime}$ region located on the same detector. 
For the second epoch, since the source was brighter, we used a larger source radius of $r_{\rm src}=70^{\prime\prime}$, and a smaller background radius of $r_{\rm bkg}=65^{\prime\prime}$.

All spectra were binned first with \texttt{ftgrouppha} using the optimal binning scheme developed by \citet{Kaastra2016}, and then further binned to have at least 20 counts per bin.

\subsection{VLA} \label{subsubsec:vla}
\begin{figure}[htbp!]
    \centering
    \includegraphics[width=\columnwidth]{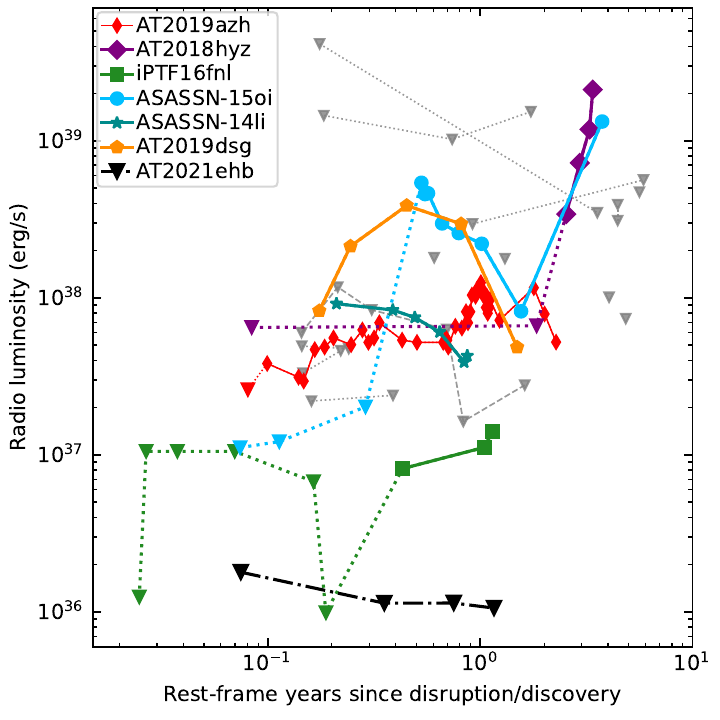}
    \caption{Radio upper limits for AT2021ehb in the context of other UV- and optically-discovered TDEs with radio data, including ASASSN-14li \citep{Alexander2016}, ASASSN-15oi \citep{Horesh2021_15oi}, iPTF16fnl \citep{Horesh2021_16fnl}, AT2018fyk \citep{Wevers2019_18fyk, Wevers2021}, AT2018hyz \citep{Cendes2022}, AT2019azh \citep{Goodwin2022, Sfaradi2022}, AT2019dsg \citep{Cendes2021_19dsg}, 
    and upper limits listed in Tab.~2 of \citet{Alexander2020}.
    \label{fig:radio_lc}}
\end{figure}

\begin{deluxetable}{ccccc}[htbp!]
	\tablecaption{Radio observations of \target.\label{tab:vla}}
	\tablehead{
		\colhead{Date}   
		& \colhead{$\Delta t$ }
		& \colhead{$\nu$} 
		& \colhead{$f_\nu$}
		& \colhead{$\nu L_{\nu}$}\\
		\colhead{}   
		& \colhead{(days)}
		& \colhead{(GHz)}
		& \colhead{($\mu$Jy)}
		& \colhead{($10^{36}\,\rm erg\,s^{-1}$)}
	}
	\startdata
2021 Mar 28.85 & $-18.8$ & 15.0 & $<16$ & $<1.8$\\ 
2021 Jul 10.53 & $83.0$ & 10.0 & $<16$ & $<1.1$\\ 
2021 Dec 5.09 & $228.0$ & 10.0 & $<16$ & $<1.1$\\ 
2022 May 6.96 & $378.1$ & 10.0 & $<14$ & $<1.1$\\ 
\enddata
\end{deluxetable}

We began a monitoring program of \target using the Very Large Array (VLA; \citealt{Perley2011}) under Program 20B-377 (PI Alexander). All of the data were analyzed following standard radio continuum image analysis procedures in the Common Astronomy Software Applications (\texttt{CASA}; \citealt{McMullin2007}). The first three observations used a custom data reduction pipeline ({\tt pwkit}; \citealt{Williams2017}), while the final observation used the standard NRAO pipeline. \target was not detected in any of our observations. All data were imaged using the CASA task {\tt clean}. We computed $3\sigma$ upper limits using the {\tt stats} command within the {\tt imtool} package of {\tt pwkit}. The result of the first epoch was reported in \citet{Alexander2021}. The full results are presented in Table~\ref{tab:vla}. 

In Figure~\ref{fig:radio_lc}, we compare the radio luminosity of \target with other UV- and optically-selected TDEs. We note that AT2021ehb looks to be significantly (by more than an order of magnitude) radio-underluminous compared to previously observed non-jetted TDEs at similar times post-peak. It has the deepest limits on any TDE radio emission at $>150$\,days post-discovery.

\section{Host Galaxy Analysis} \label{sec:host}

\begin{figure}[htbp!]
    \centering
    \includegraphics[width=0.7\columnwidth]{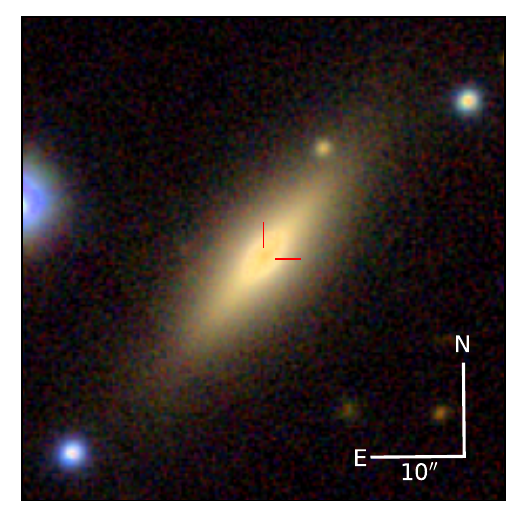}
    \caption{PS1 RGB false-color $g$/$i$/$z$ image centered on \target.
    North is up and east to the left. A 10$^{\prime\prime}$ scale bar is included.
    \label{fig:host}}
\end{figure}

Figure~\ref{fig:host} shows the pre-TDE optical image centered on \target, using data from PS1. The host galaxy appears to be close to edge-on. 

\subsection{Velocity Dispersion and Black Hole Mass} \label{subsec:esi}
The host galaxy absorption lines are prominent in the optical spectra (see Figure~\ref{fig:opt_spec}). Using our medium-resolution ($R=5350$) spectrum taken with Keck-II/ESI, we measured the line centers of strong absorption lines, and determined the redshift to be $z = 0.0180$. 

Following previous TDE work \citep{Wevers2017, Wevers2019_Mbh, French2020}, we measured the stellar velocity dispersion by fitting the normalized ESI spectrum (see pre-processing procedures in Appendix~\ref{sec:optspec_details}) with the penalized pixel-fitting (\texttt{pPXF}) software \citep{Cappellari2004, Cappellari2017}. \texttt{pPXF} fits the absorption line spectrum by convolving a library of stellar spectra with Gauss-Hermite functions. We adopted the ELODIE v3.1 high resolution ($R=42000$) template library \citep{Prugniel2001, Prugniel2007}. 

To robustly measure the velocity dispersion and the associated uncertainties, we performed 1000 Monte Carlo (MC) simulations, following the approach adopted by \citet{Wevers2017}. In each fitting routine, we masked wavelength ranges of common galaxy emission lines and hydrogen Balmer lines. 
The derived velocity dispersion is $\sigma = 92.9^{+5.3}_{-5.2}\,{\rm km\,s^{-1}}$ at 95\% confidence interval. 

According to the $M_{\rm BH}$--$\sigma$ relation \citep{Kormendy2013}, the measured $\sigma$ corresponds to a black hole mass of
log$(M_{\rm BH}/M_\odot) = 7.03\pm 0.15\,({\rm stat}) \pm 0.29\,({\rm sys})$,
where 0.29 is the intrinsic scatter of the $M_{\rm BH}$--$\sigma$ relation. 
If adopting the \cite{Ferrarese2005} $M_{\rm BH}$--$\sigma$ relation, then
log$(M_{\rm BH}/M_\odot) = 6.60\pm 0.20 \,({\rm stat})\pm 0.34\,({\rm sys})$.
Hereafter we adopt the result from the \citet{Kormendy2013} relation because it includes more low-mass galaxies.

We note that although the \citet{Kormendy2013} relation was originally calibrated mainly at a $M_{\rm BH}$ regime that is too massive to produce a TDE, recent studies show that the same relation holds in the dwarf galaxy regime \citep{Baldassare2020}. 

\subsection{Host SED Model} \label{subsec:host_fsps}

\begin{figure}[htbp!]
    \centering
    \includegraphics[width=\columnwidth]{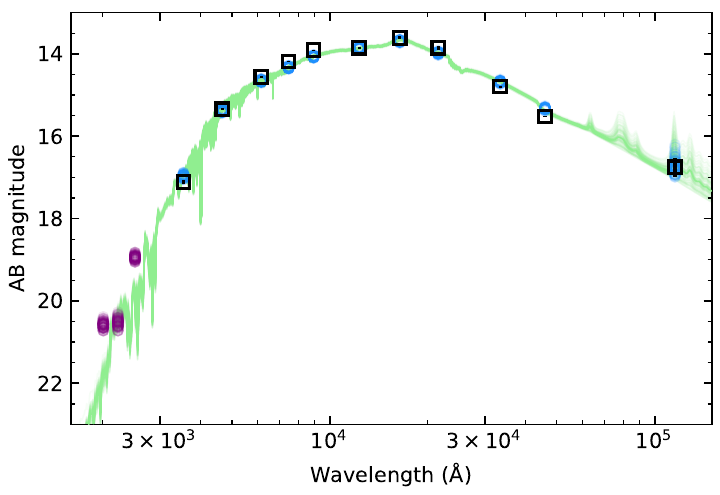}
    \caption{Host galaxy SED of \target. 
    The open squares are the Galactic extinction-corrected host photometry (see Table~\ref{tab:host}). 
    The green lines are samples from the posterior distribution of host galaxy SED models. 
    The open circles are the synthetic host galaxy magnitudes in the observed bands (shown in blue) and in the UV filters of \swift/UVOT (shown in purple).
    \label{fig:host_sed}}
\end{figure}

We constructed the pre-TDE host galaxy SED using photometry from SDSS, the Two Micron All-Sky Survey (2MASS; \citealt{Skrutskie2006}), and the AllWISE catalog \citep{Cutri2014}. The photometry of the host is shown in Table~\ref{tab:host}.

\begin{deluxetable}{cccc}[htbp!]
    \tablecaption{Observed photometry of the host galaxy.\label{tab:host}}
	\tablehead{
	\colhead{Catalog}
		& \colhead{Band} 
		& \colhead{$\lambda_{\rm eff}$ (nm)}
		& \colhead{Magnitude} 
	}
	\startdata
SDSS & $u$ & 355 & $17.748 \pm 0.019$ \\
SDSS & $g$ & 467 & $15.814 \pm 0.003$ \\
SDSS & $r$ & 616 & $14.901 \pm 0.002 $ \\
SDSS & $i$ & 747 & $14.443 \pm 0.003 $ \\
SDSS & $z$ & 892 & $14.094 \pm 0.004 $ \\
2MASS & $J$ & 1232 & $13.951 \pm 0.025 $ \\
2MASS & $H$ & 1642 & $13.676 \pm 0.034 $ \\
2MASS & $K_{\rm s}$ & 2157 & $13.893 \pm 0.043 $ \\
AllWISE & $W1$ & 3346 & $14.816 \pm 0.024$ \\
AllWISE & $W2$ & 4595 & $15.535 \pm 0.022$ \\
AllWISE & $W3$ & 11553 & $16.756 \pm 0.229$ \\
	\enddata
\end{deluxetable}

Our SED fitting approach is similar to that described in \citet{vanVelzen2021}. We used the flexible stellar population synthesis (\texttt{FSPS}) code \citep{Conroy2009}, and adopted a delayed exponentially declining star-formation history (SFH) characterized by the $e$-folding timescale $\tau_{\rm SFH}$. The \texttt{Prospector} package \citep{Johnson2021} was utilized to run a Markov Chain Monte Carlo sampler \citep{Foreman-Mackey2013}. We show the best-fit model prediction of the host galaxy optical spectrum at the bottom of Figure~\ref{fig:opt_spec}.

From the marginalized posterior probability functions we obtain
the total galaxy stellar mass log$(M_{\ast}/M_\odot) = 10.18_{-0.02}^{+0.01}$, 
the metallicity, ${\rm log}Z = -0.57\pm0.04$, 
$\tau_{\rm SFH} = 0.19_{-0.07}^{+0.18}$\,Gyr, 
the population age, $t_{\rm age} = 12.1_{-0.6}^{+0.3}$\,Gyr, 
and negligible host reddening ($E_{B-V, \rm host} = 0.01\pm0.01$\,mag). 
The best-fit SED model is shown in Figure~\ref{fig:host_sed}.

Following \citet{Gezari2021}, we use the $M_{\rm BH}$--$M_{\ast}$ relation from \citet{Greene2020} to obtain a black hole mass of log$(M_{\rm BH}/M_\odot) = 7.14\pm(0.10 + 0.79)$, where 0.79 is the intrinsic scatter of the scaling relation. 
This is consistent with the $M_{\rm BH}$ inferred from the $M_{\rm BH}$--$\sigma$ relation (\S\ref{subsec:esi}). 

To summarize, the host galaxy of \target has a total stellar mass of $M_\ast \approx 10^{10.18}\,M_\odot$ and a BH mass of $M_{\rm BH}\approx 10^{7.03}\,M_\odot$.
The measured black hole mass is on the high end of the population of optically selected TDEs \citep{French2020, Nicholl2022}, and is too massive to disrupt a white dwarf \citep{Rosswog2009}. 


\section{Analysis of the TDE Emission} \label{sec:TDEanalysis}

\subsection{UV/optical Photometric Analysis} \label{subsec:uvopt_bbfit}
\begin{figure}[htbp!]
    \centering
    \includegraphics[width=\columnwidth]{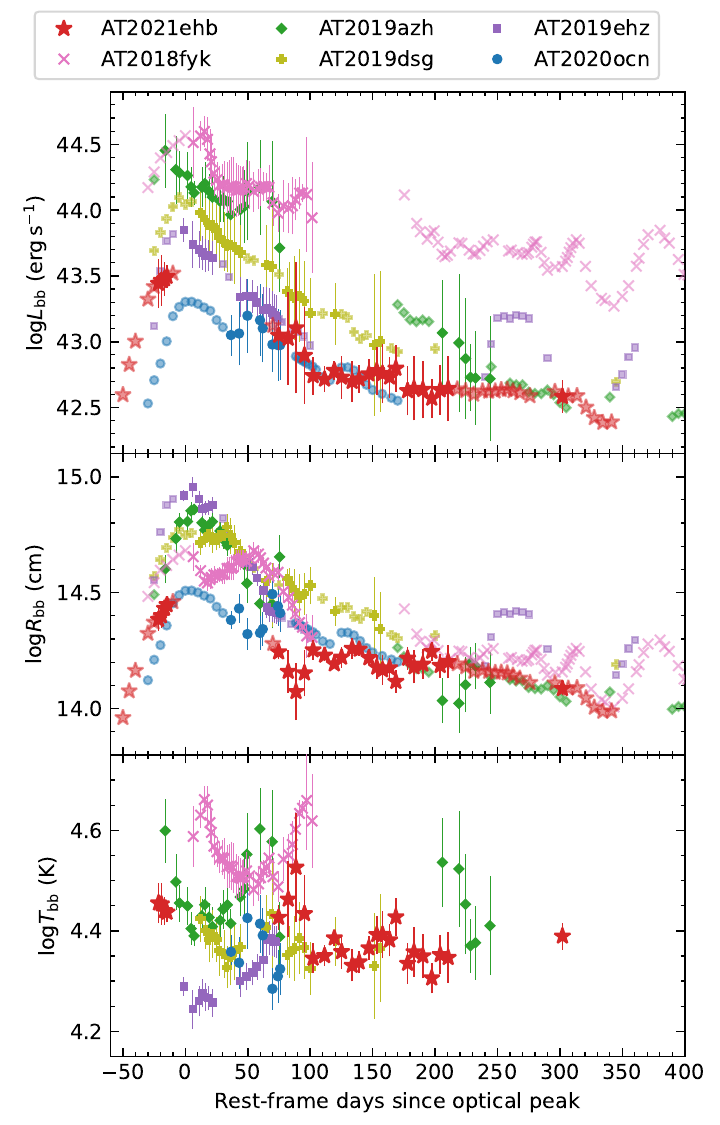}
    \caption{Evolution of the UV/optical blackbody properties of AT2021ehb compared with a sample of recent X-ray bright TDEs in the literature, including AT2018fyk \citep{Wevers2019_18fyk, Wevers2021}, AT2019dsg \citep{Stein2021}, AT2019azh \citep{Hinkle2021}, AT2020ocn, and AT2019ehz \citep{vanVelzen2021}. The results of ``good epochs'' (see definition in the text) are shown in high-opacity colors, whereas results of ``ok epochs'' are shown in semi-transparent. \label{fig:bbfit}}
\end{figure}

To capture the general trend of \target's UV/optical photometric evolution, we fit the data in each filter using a combination of five-order polynomial functions and Gaussian process smoothing, following procedures described in Appendix B.4 of \citet{Yao2020}. 
The model fits in $r_{\rm ZTF}$, $uvw1$, and $uvw2$ are shown as semi-transparent lines in Figure~\ref{fig:lc_uvopt}.

We then define a set of ``good epochs'' close in time to actual multiband measurements, and fit a Planck function to each set of fluxes to determine the effective temperature $T_{\rm bb}$, photospheric radius $R_{\rm bb}$, and blackbody luminosity of the UV/optical emitting component $L_{\rm bb}$. We initially assume $E_{B-V, \rm host}=0$\,mag, and then repeat the procedure under different assumptions about the host reddening. We find that the fitting residual monotonically increases as $E_{B-V, \rm host}$ increases from 0\,mag to 0.2\,mag, suggesting negligible host reddening. Therefore, for the reminder of the discussion we assume $E_{B-V, \rm host}=0$\,mag.

We also define a set of ``ok epochs'' where we only have photometric observations in the optical (or only in the UV). Due to a lack of wavelength coverage, $T_{\rm bb}$ and $R_{\rm bb}$ can not be simultaneously constrained. As such we fix the $T_{\rm bb}$ values by interpolating the $T_{\rm bb}$ evolution of ``good epochs'', and fit for $R_{\rm bb}$ values of ``ok epochs''.

The physical parameters derived from the blackbody fits are \ad{presented in Table~\ref{tab:bbpars} (Appendix~\ref{subsec:modelfit}) and} shown in Figure~\ref{fig:bbfit}, where they are compared with a sample of recent TDEs with multiple X-ray detections. We have measured the blackbody parameters of other TDEs using the same procedures described above.

While the temperature of AT2021ehb ($T_{\rm bb}\sim 2.5\times10^4$\,K) is typical among optical and X-ray bright TDEs, its peak radius ($R_{\rm bb}\sim 3\times 10^{14}\,{\rm cm}$) and luminosity ($L_{\rm bb}\sim 3\times 10^{43}\,{\rm erg\,s^{-1}}$) are at the low end of the distributions. We note that in the ZTF-I sample of 30 TDEs \citep{Hammerstein2022}, only two objects (AT2020ocn and AT2019wey) have peak radii smaller than that of AT2021ehb (see the discussion in \S\ref{subsubsec:whynoline}). 

\subsection{Optical Spectral Analysis} \label{subsec:opt_spec_analyze}

\begin{figure}[htbp!]
    \centering
    \includegraphics[width=\columnwidth]{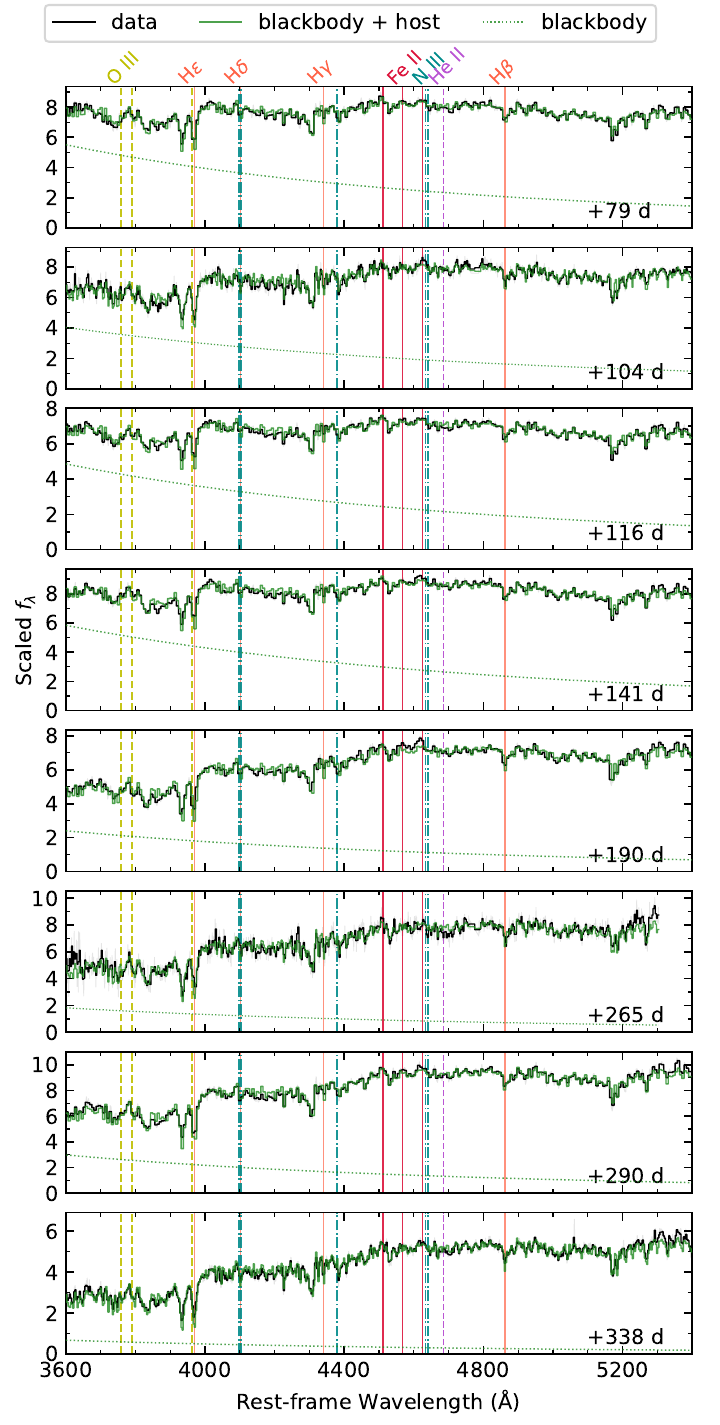}
    \caption{Long-slit optical spectra of \target taken at 8 different epochs. 
    The spectrum ($f_{\rm \lambda, obs}$) is plotted in black.
    The blackbody continuum ($A_1 f_{\rm \lambda, BB}$; dotted lines) plus host galaxy spectrum ($A_2 f_{\rm \lambda, host}$) is plotted in green. 
    No spectral features commonly seen in optically selected TDEs are observed in \target.
    \label{fig:spec_model}}
\end{figure}

\begin{figure*}
\centering
    \includegraphics[width=0.92\textwidth]{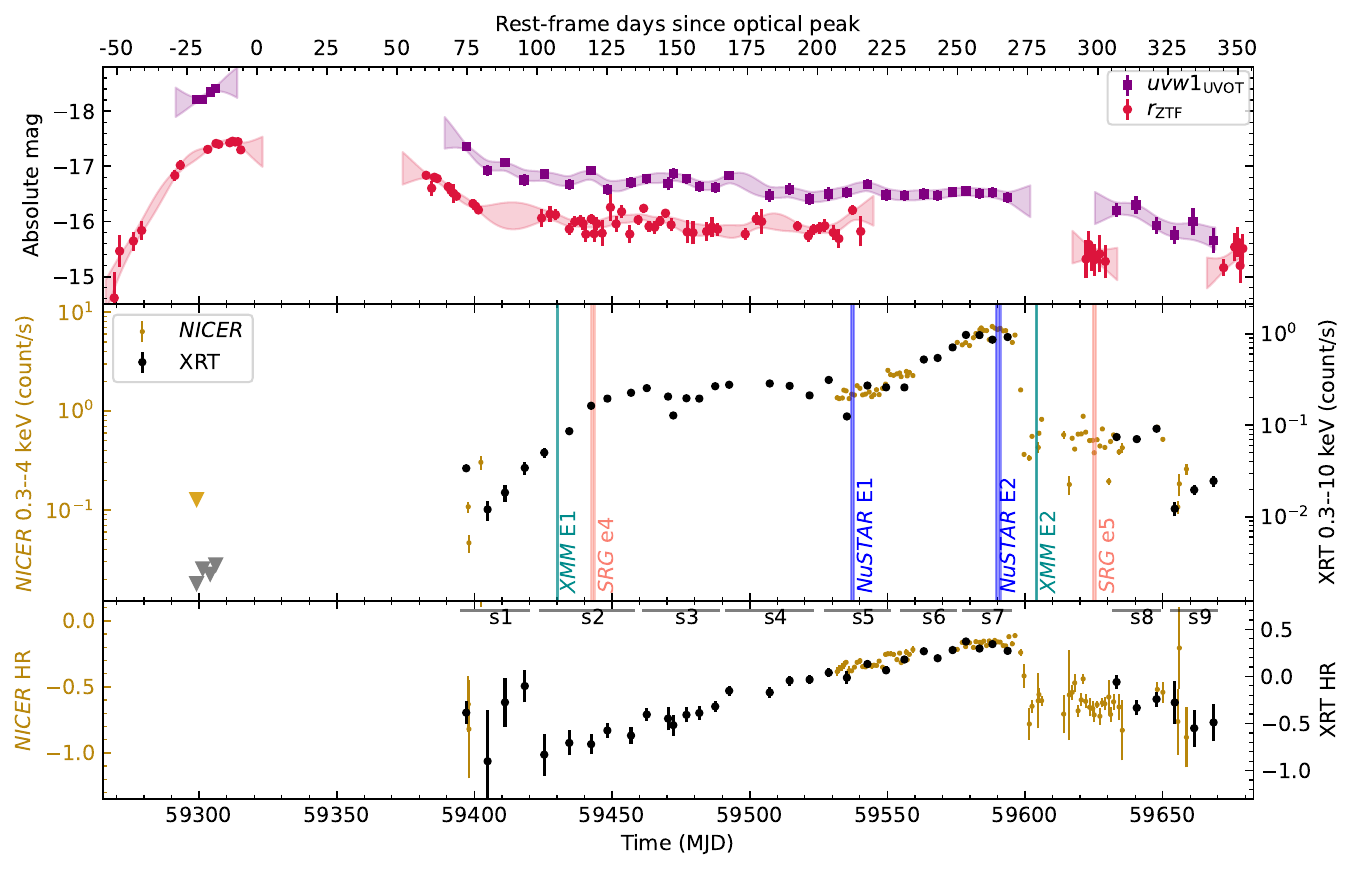}
    \caption{\textit{Upper}: UV ($uvw1_{\rm UVOT}$) and optical ($r_{\rm ZTF}$) light curves of \target. 
    \textit{Middle}: XRT and \nicer X-ray net count rates of \target. Epochs of \xmm, \srg, and \nustar observations are marked by the vertical lines.
    \textit{Lower}: XRT and \nicer hardness ratio (HR) evolution of \target.
    \label{fig:xray_lc}}
\end{figure*}

Figure~\ref{fig:opt_spec} shows that no broad line is evident in the optical spectra of \target. To search for weak spectral features from the TDE, we fit the Galactic extinction-corrected long-slit spectra in rest-frame 3600--5400\,\AA\ using a combination of blackbody emission and host galaxy contribution: $f_{\rm \lambda, obs} = A_1 f_{\rm \lambda, BB} + A_2 f_{\rm \lambda, host}$. 
Here $f_{\rm \lambda, BB} = \pi B_{\lambda}(T_{\rm bb}) (R_{\rm bb}^2/D_{L}^2)$, where $T_{\rm bb}$ and $R_{\rm bb}$ are obtained by linearly interpolating the blackbody parameters derived in \S\ref{subsec:uvopt_bbfit} at the relevant $\delta t$.
$f_{\rm \lambda, host}$ is the predicted host galaxy spectrum obtained in \S\ref{subsec:host_fsps} convolved with the instrumental broadening $\sigma_{\rm inst}$ (see Appendix~\ref{sec:optspec_details}).
$A_1$ and $A_2$ are constants added to account for unknown factors, including the varying amount of host galaxy flux falling within the slit (which depends on the slit width, slit orientation, seeing conditions and target acquisition), uncertainties in the absolute flux calibration and the adopted blackbody parameters. 
We note that $f_{\rm \lambda, host}$ is the predicted spectrum for the whole galaxy, and therefore might not be a perfect description of the bulge spectrum.

The fitting results are shown in Figure~\ref{fig:spec_model}. 
We mark locations of emission lines commonly seen in TDEs, including Balmer lines, \ion{He}{II}, the Bowen fluorescence lines of \ion{N}{III} and \ion{O}{III}, as well as low-ionization \ion{Fe}{II} lines 
\citep{Blanchard2017, Wevers2019_18fyk, vanVelzen2021}. 
The observed spectra of \target can be well described by a blackbody continuum (dotted lines) plus host galaxy contribution.
The spectra at $\delta t > 170$\,days are mostly from the host, and therefore it is not very surprising that no discernible TDE lines were detected. 
However, at $\delta t<170$\,days, the blackbody component contributes 25\%--80\% of the total flux. 
As such, it is surprising that no prominent lines from the TDE itself can be identified. 
We further discuss this result in \S\ref{subsubsec:whynoline}.

\subsection{X-ray Light Curve Analysis}

The middle panel of Figure~\ref{fig:xray_lc} shows the XRT and \nicer (all binned by obsID) light curves. 
The lower panel of Figure~\ref{fig:xray_lc} shows the evolution of the hardness ratio, defined as ${\rm HR} \equiv (H-S)/(H+S)$, where $H$ is the number of net counts in the hard band, and $S$ is the number of net counts in 0.3--1\,keV. For XRT we take 1--10\,keV as the hard band, while for \nicer we take 1--4\,keV.

X-rays were not detected at $\delta t < 0$. Pre-peak X-ray upper limits are provided by \swift/XRT ($<10^{40.9}\,{\rm erg\,s^{-1}}$, Table~\ref{tab:xrt}) and \srg/eROSITA ($<10^{40.2}\,{\rm erg\,s^{-1}}$, Table~\ref{tab:srg}). 

X-rays were first detected by XRT at $\delta t = 73.9$\,days. 
The exact time of the X-ray onset cannot be accurately constrained. The count rate initially exhibited strong variability from $\delta t = 73.9$\,days to $\delta t = 82.3$\,days, and then gradually increased out to $\delta t=250$\,days. 
At the same time, the HR gradually increased. 
From $\delta t=250$\,days to $\delta t=271$\,days, both the X-ray flux and the hardness stayed at the maximum values. 

From $\delta t = 271.0$\,days to $\delta t=273.7$\,days, the \nicer net count rate suddenly decreased by a factor of 10 \citep{Yao2022_ATel15217}. At the same time, the HR significantly decreased. After an X-ray plateau of $\approx 50$\,days, the XRT net count rate further decreased drastically by a factor of 6 (from $\delta t = 320.9$\,days to $\delta t=327.2$\,days).

\subsection{X-ray Spectral Analysis}

In this subsection, we first present a joint spectral analysis of contemporaneous data sets obtained from \nicer and \nustar, including the first epoch in 2021 November 18--19 (\S\ref{subsubsec:joint1}) and the second epoch in 2022 January 10--12 (\S\ref{subsubsec:joint2}). These observations are of high SNR and cover a wide energy range. As such, the fitting results can guide us to choose appropriate spectral models to fit spectra with lower SNR.
We then perform analysis on data sets obtained by single telescopes, including \xmm (\S\ref{subsubsec:XMMspec}), \srg (\S\ref{subsubsec:SRGspec}), \swift/XRT (\S\ref{subsubsec:XRTspec}),   and \nicer (\S\ref{subsubsec:NICERspec}).

All spectral fitting was performed with \texttt{xspec} (v12.12, \citealt{Arnaud1996}). 
We used the \texttt{vern} cross sections \citep{Verner1996}. The \texttt{wilm} abundances \citep{Wilms2000} were adopted in \S\ref{subsubsec:joint1} and \S\ref{subsubsec:joint2}, whereas the \citet{Anders1989} abundances were adopted in \S\ref{subsubsec:XMMspec}, \S\ref{subsubsec:SRGspec}, \S\ref{subsubsec:XRTspec}, and \S\ref{subsubsec:NICERspec}.

\subsubsection{\textit{NICER}+\nustar First Epoch, 2021 November} \label{subsubsec:joint1}

\begin{figure}[htbp!]
    \centering
    \includegraphics[width=\columnwidth]{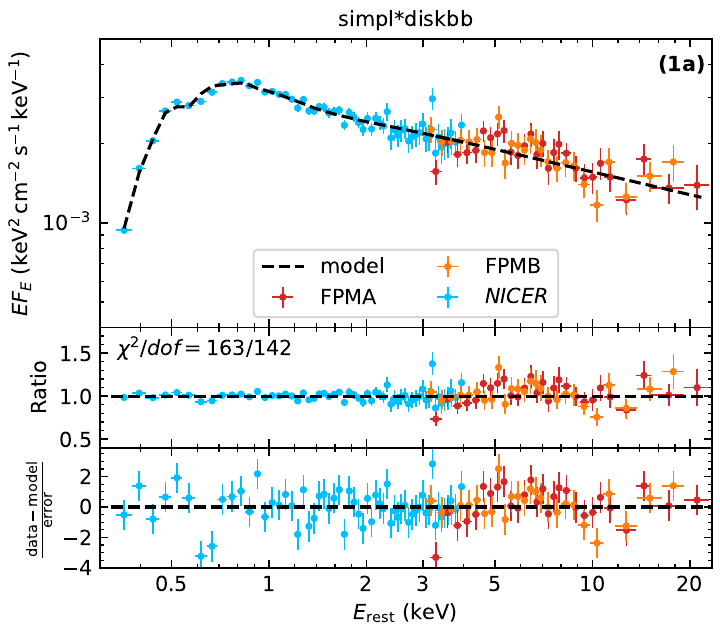}
    \caption{The spectrum of the first joint \nicer and \nustar observations (2021 November). 
    See Table~\ref{tab:joint1_pars} for best-fit parameters. 
    \nustar/FPMB and \nicer data have been divided by $C_{\rm FPMB}$ and $C_{\rm NICER}$, respectively.
    The data have been rebinned for visual clarity.
    }
    \label{fig:joint1_mods}
\end{figure}

We chose energy ranges where the source spectrum dominates over the background. 
For \nicer we used 0.3--4\,keV. 
For \nustar/FPMA we used 3--23\,keV, and for FPMB we used 3--20 keV\footnote{In this \nustar observation, FPMB is more affected by a nearby bright source.}.
All data were fitted using $\chi^2$-statistics.

\begin{deluxetable}{llc}
\tablecaption{Modeling of the first joint \nicer and \nustar observations, $\delta t=212$\,days.
\label{tab:joint1_pars} }
\tablehead{
\colhead{Component}
& \colhead{Parameter}
& \colhead{(1a)}
}
\startdata
\texttt{constant} & $C_{\rm FPMB}$ 
& $1.03_{-0.05}^{+0.06}$ 
\\
& $C_{\rm NICER}$ 
& $0.85_{-0.05}^{+0.06}$ 
\\
\texttt{ztbabs} &  $N_{\rm H}$ ($10^{20}\,{\rm cm^{-2}}$)
& $<0.75$ 
\\
\texttt{simpl} & $\Gamma$
& $2.29 \pm 0.05$  
\\
& $f_{\rm sc}$
& $0.35_{-0.03}^{+0.02}$  
\\
\texttt{diskbb} & $T_{\rm in}$ (eV) 
& $164_{-9}^{+6}$  
\\
& $R_{\rm in}^\ast$ ($10^4\,{\rm km}$) 
& $25.5^{+4.4}_{-2.0}$  
\\
--- & $\chi^2/dof$
& $163.17/142=1.15$ 
\\
\enddata
\end{deluxetable}

For all spectral models described below, we included the Galactic absorption using the \texttt{tbabs} model \citep{Wilms2000}, with the hydrogen-equivalent column density $N_{\rm H}$ fixed at $9.97\times 10^{20}\,{\rm cm^{-2}}$ \citep{HI4PI2016}. 
We shifted the TDE emission using the convolution model \texttt{zashift}, with the redshift $z$ fixed at $0.018$.
We included possible absorption intrinsic to the source using the \texttt{ztbabs} model.
We also included a calibration coefficient (\texttt{constant}; \citealt{Madsen2017}) between FPMA, FPMB, and \nicer, with $C_{\rm FPMA}\equiv 1$. This term also accounts for the differences in the mean flux between \nustar and \nicer that results from intrinsic source variability.

First, we fitted the spectrum with a power-law (PL), and obtained a photon index of $\Gamma \approx 2.7$. 
The fit is unacceptable, with the reduced $\chi^2$ being $\chi^2_{\rm r} = 3.44$ for 144 degrees of freedom ($dof$). 
The residuals are most significant between 0.3 and 2\,keV, suggesting the existence of a (thermal) soft component.
Therefore, we changed the PL to \texttt{simpl*thermal\_model}. 
Here \texttt{simpl} is a Comptonization model that generates the PL component via Compton scattering of a fraction ($f_{\rm sc}$) of input seed photons \citep{Steiner2009}. The flag $R_{\rm up}$ was set to 1 to only include upscattering. 
We experimented with three different thermal models: a blackbody (\texttt{bbody}), a multicolor disk (MCD; \texttt{diskbb}; \citealt{Mitsuda1984}), and a single-temperature thermal plasma (\texttt{bremss}; \citealt{Kellogg1975}), resulting in $\chi^2_{\rm r}=1.33$, 1.15, and 1.35 (for $dof=142$), respectively. 
The fit statistics favors a MCD.

The best-fit result with a MCD, defined as model (1a), is shown in Figure~\ref{fig:joint1_mods}. 
We present the best-fit parameters in Table~\ref{tab:joint1_pars}. 
Here $T_{\rm in}$ is the inner disk temperature, and $R_{\rm in}^{\ast} \equiv R_{\rm in} \sqrt{\cos i}$ is the apparent inner disk radius times square root of $\cos i$, where $i$ is the system inclination. 
$R_{\rm in}^{\ast}$ is inferred from the normalization parameter of \texttt{diskbb}. 
Model (1a) gives a good fit with $\chi^2_{\rm r} = 163/142=1.15$. 

\begin{deluxetable*}{ll|cccc}
\tablecaption{Modeling of the second joint \nicer and \nustar observations, $\delta t=264$\,days. 
\label{tab:joint2_pars} }
\tablehead{
\colhead{Component }
& \colhead{Parameter}
& \colhead{(2a)}
& \colhead{(2b)}
& \colhead{(2c)}
& \colhead{(2d)}
}
\startdata
\texttt{constant} & $C_{\rm FPMB}$ 
& $1.03$ 
& $1.03 \pm 0.01$ 
& $1.03 \pm 0.01$ 
& $1.03 \pm 0.01$   
\\
& $C_{\rm NICER}$ 
& $0.98$ 
& $1.02 \pm 0.01$ 
& $1.02 \pm 0.01$ 
& $1.03\pm 0.01$  
\\
\texttt{tbfeo} & $A_{\rm O}$
& 1.40 
& $1.48^{+0.11}_{-0.08}$ 
& $1.26 \pm 0.13$ 
& $1.45^{+0.10}_{-0.07}$ 
\\
& $A_{\rm Fe}$
& 1.80  
& $2.07^{+0.63}_{-0.64}$ 
& $1.99_{-0.62}^{+0.61}$  
& $2.37^{+0.50}_{-0.39}$   
\\
\texttt{ztbabs} &  $N_{\rm H}$ ($10^{20}\,{\rm cm^{-2}}$)
& 0.00 
& $<0.12$ 
& $0.70\pm0.28$ 
& $<0.01$ 
\\
\texttt{diskbb} & $T_{\rm in}$ (eV) 
& 187 
& $198^{+8}_{-6}$ 
& $257\pm8$  
& $180^{+7}_{-2}$ 
\\
& $R_{\rm in}^\ast$ ($10^4\,{\rm km}$) 
& 31.7  
& $28.4_{-1.7}^{+1.2}$ 
& $10.5_{-0.9}^{+1.0}$  
& $47.3\pm 2.8$  
\\
\texttt{simpl} & $\Gamma$
& 2.09   
& $2.11 \pm 0.01$ 
& ...
& $2.26 \pm 0.01$   
\\
& $f_{\rm sc}$
& 0.52  
& $0.49^{+0.02}_{-0.03}$ 
& ...
& $0.61 \pm 0.01$   
\\
\texttt{gaussian} & $E_{\rm line}$ (keV)
& ... 
& $4.92_{-0.71}^{+0.36}$
& ... 
& ... 
\\
& $\sigma_{\rm line}$ (keV)
& ... 
& $2.18_{-0.32}^{+0.50}$
& ... 
& ...
\\
& Norm ($10^{-4}\,{\rm ph\,cm^{-2}\,s^{-1}}$)
& ... 
& $2.52^{+1.01}_{-0.51}$
& ... 
& ...
\\
\texttt{relxill} & $q_1 = q_2$ 
& ... 
& ... 
& $3$ (frozen)    
& ... 
\\
& $a$
& ...
& ... 
& 0 (frozen)     
& ... 
\\
& $i$ ($^{\circ}$)
& ...
& ... 
& $43.4^{+8.5}_{-9.6}$     
& ... 
\\
& $R_{\rm in}$ ($R_{\rm ISCO}$)
& ...
& ... 
& $1$ (frozen)     
& ... 
\\
& $R_{\rm out}$ ($R_{\rm g}$)
& ...
& ... 
& $400$ (frozen)     
& ... 
\\
& $\Gamma$
& ...
& ... 
& $1.86\pm0.02$     
& ... 
\\
& log$\xi$ ($\rm erg\,cm\,s^{-1}$)
& ...
& ... 
& $4.09_{-0.12}^{+0.20}$     
& ... 
\\
& $A_{\rm Fe}$
& ...
& ... 
& $1.86_{-0.63}^{+1.46}$     
& ... 
\\
& $E_{\rm cut}$ (keV)
& ...
& ... 
& $54.0_{-9.5}^{+13.4}$     
& ... 
\\
& $R_{\rm F}$
& ...
& ... 
& 1 (frozen)     
& ... 
\\
& Norm ($10^{-5}$)
& ...
& ... 
& $6.1^{+0.40}_{-0.40}$     
& ... 
\\
\texttt{xstar} & $N_{\rm H}$ ($10^{23}\,{\rm cm^{-2}}$) 
& ... 
& ...
& ... 
& $2.22^{+0.49}_{-0.86}$ 
\\
& ${\rm log}\xi$ ($\rm erg\,cm\,s^{-1}$)
& ... 
& ... 
& ... 
& $1.51^{+0.34}_{-0.32}$ 
\\
& $f_{\rm cover}$
& ... 
& ... 
& ... 
& $0.31\pm 0.02$ 
\\
& Redshift
& ... 
& ... 
& ... 
& 0 (frozen)
\\
--- & $\chi^2/dof$
& $609.43/299=2.04$     
& $330.72/296=1.12$     
& $306.64/295=1.04$     
& $318.23/296=1.08$     
\\
\enddata
\tablecomments{Parameter uncertainties of model (2a) cannot be calculated since $\chi^2_{\rm r}>2$. }
\end{deluxetable*}

\subsubsection{\textit{NICER}+\nustar Second Epoch, 2022 January} \label{subsubsec:joint2}

We chose energy ranges where the source spectrum dominates over the background. 
For \nicer we used 0.3--7.0\,keV; For \nustar FPMA and FPMB we used 3--30\,keV.
All data were fitted using $\chi^2$-statistics. 
Unlike in \S\ref{subsubsec:joint1}, here we use \texttt{tbfeo} to model the Galactic absorption. 
Compared with \texttt{tbabs}, \texttt{tbfeo} allows the O and Fe abundances ($A_{\rm O}$, $A_{\rm Fe}$) to be free.

\begin{figure}[htbp!]
    \centering
    \includegraphics[width=\columnwidth]{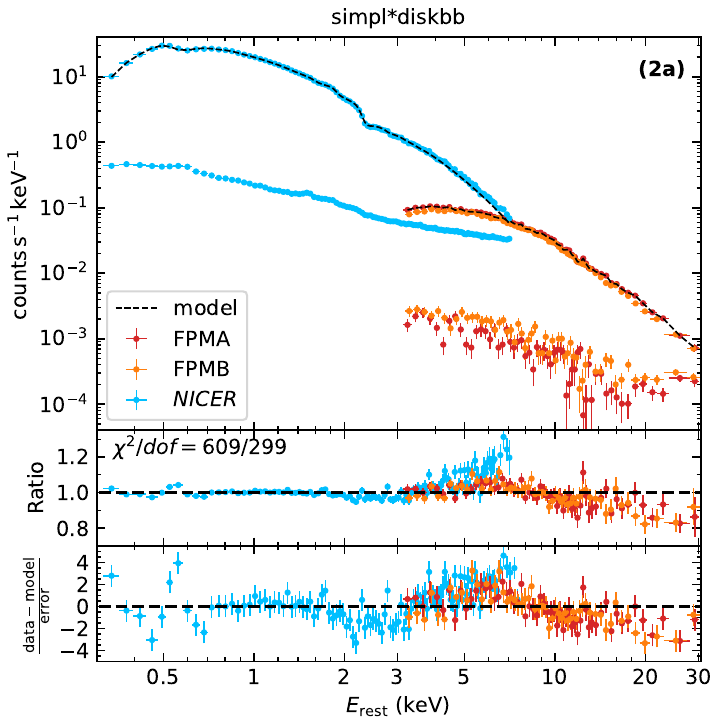}
    \caption{The spectrum of the second joint \nicer and \nustar observations (2022 January). This figure highlights the flux levels of the source and background spectra.
    }
    \label{fig:joint2_initial}
\end{figure}

\begin{figure*}[htbp!]
    \centering
    \includegraphics[width=\columnwidth]{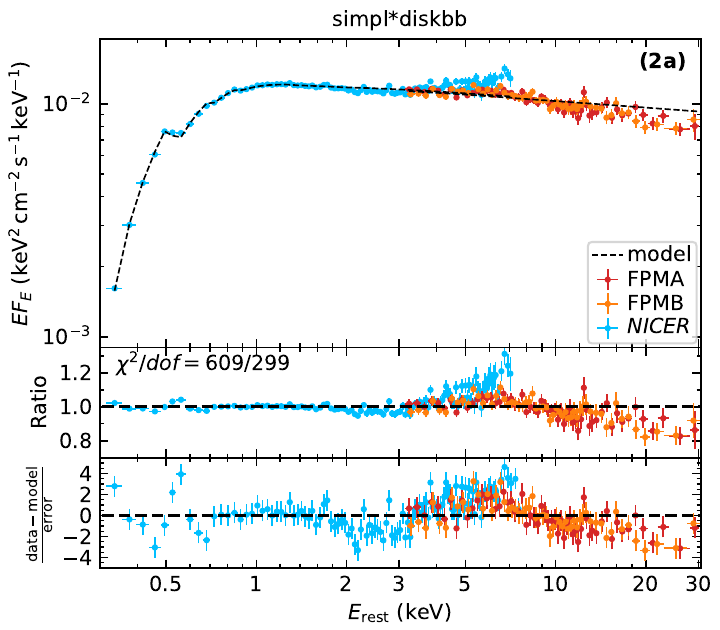}
    \includegraphics[width=\columnwidth]{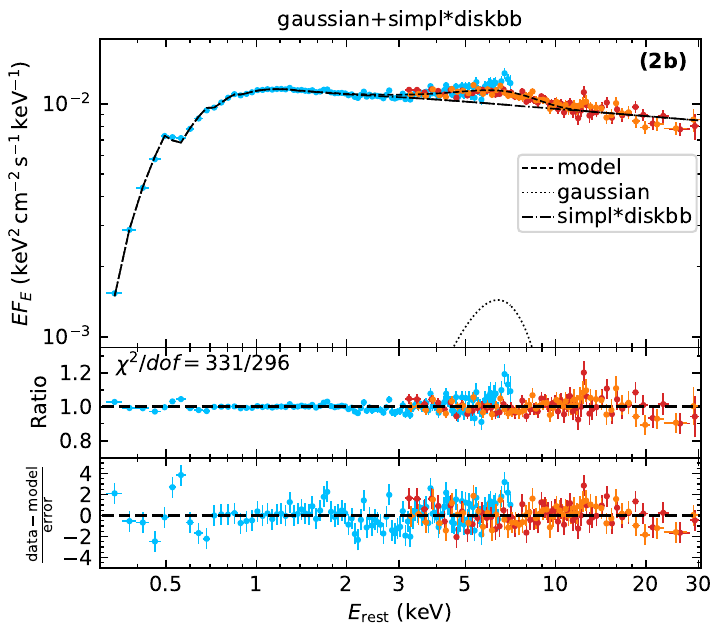}\\
    \includegraphics[width=\columnwidth]{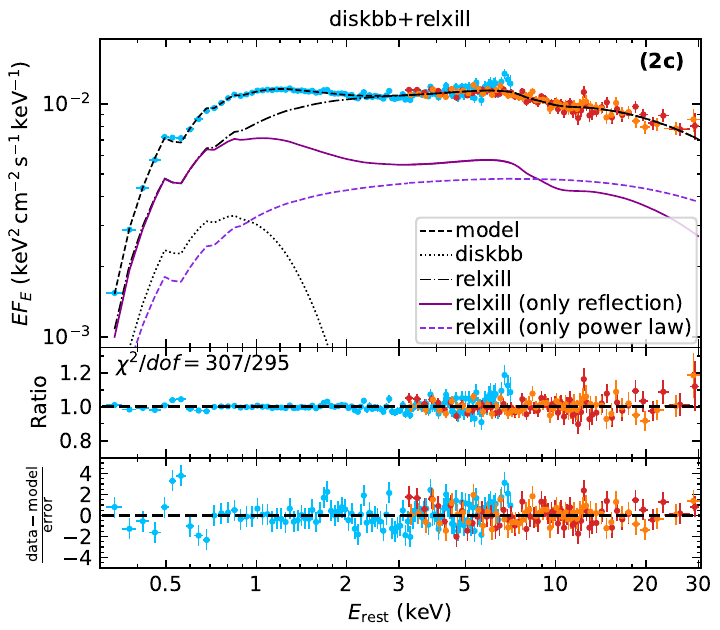}
    \includegraphics[width=\columnwidth]{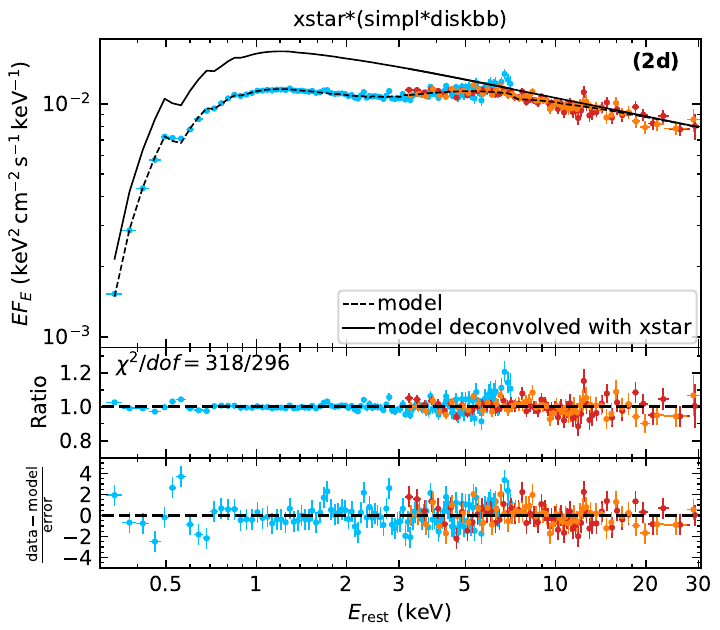}
    \caption{The spectrum of the second joint \nicer and \nustar observations (2022 January). 
    See Table~\ref{tab:joint2_pars} for best-fit parameters. 
    \nustar/FPMB and \nicer data have been divided by $C_{\rm FPMB}$ and $C_{\rm NICER}$, respectively.
    The data have been rebinned for visual clarity.
    In the bottom left panel, we also show the reflection contribution and the direct power law contribution individually from \texttt{relxill}. 
    In the bottom right panel, we also show the best-fit model (2d) deconvolved with the \texttt{xstar} component.
    }
    \label{fig:joint2_mods}
\end{figure*}

We adopted a continuum model of \texttt{simpl*diskbb}, defined as (2a).
The result, with $\chi^2_{\rm r}=2.04$, is shown in Figure~\ref{fig:joint2_initial} and the upper left panel of Figure~\ref{fig:joint2_mods}. 
The best-fit parameters are given in Table~\ref{tab:joint2_pars}. 
The residual plot clearly indicates the existence of unmodeled spectral features and a significant offset between \nustar and \nicer at 6--7\,keV. 

First, we study whether this offset is brought about by a cross-calibration difference between \nicer and \nustar.
To this end, we replaced \texttt{constant} with \texttt{crabcorr} \citep{Ludlam2022}, which multiplies the spectrum by a power-law of $C \cdot E^{- \Delta \Gamma}$. 
When $\Delta \Gamma =0$, \texttt{crabcorr} is equivalent to \texttt{constant}. 
We fixed $\Delta \Gamma_{\rm FPMA} = \Delta \Gamma_{\rm FPMB} = 0$, and allow $\Delta \Gamma_{\nicer}$ to be free. 
The best-fit model gives $\Delta \Gamma_{\nicer}=-0.128_{-0.023}^{+0.014}$, which is too large compared with the value of $\Delta \Gamma_{\nicer}\approx -0.06$ found by \citet{Ludlam2022}. Therefore, we conclude that a difference in the cross-calibration slope is likely not the primary reason for the 6--7\,keV offset.

Next, we investigate whether this offset is caused by imperfect \nicer calibration at 2--3\,keV. \nicer effective area changes rapidly in the 2--3\,keV band. Calibration issues in that range may cause the model to over-estimate the data at 2--3\,keV, and to badly under-estimate it above 3\,keV. As a test, we performed the fit omitting the 2--3\,keV region in the \nicer data. However, the best-fit result still leaves a significant offset between \nicer and \nustar at 6--7\,keV, similar to that shown in Figure~\ref{fig:joint2_initial}.

We are left to conclude that the offset is likely caused by either an underestimate of \nicer background at the high energy end or systematic uncertainties in \nicer calibration that is not well characterized. This conjecture is based on the fact that \nicer uses X-ray concentrators optics and its 3--7\,keV background is $>10$ times brighter than that of \nustar (Figure~\ref{fig:joint2_initial}). 
On the other hand, \nustar adopts X-ray focusing optics, which enables more robust background estimation using regions close to the object of interest. 

In the following, we attempted to improve the fit by three approaches: adding a Gaussian line, adding reflection emission features, and adding absorption features.

\paragraph{Modeling with a Gaussian Line Profile}

The result with adding a \texttt{gaussian} line component is shown in the upper-right panel of Figure~\ref{fig:joint2_mods}. This model, defined as (2b), provides a much better fit compared with (2a). 
The best-fit parameters (Table~\ref{tab:joint2_pars}) give a very broad emission profile with a central energy at $E_{\rm line}\sim 5$\,keV and a line width of $\sigma_{\rm line}\sim2$\,keV. If the 3--7\,keV spectral feature indeed comes from an emission line, its central energy is different from the emission line at $E_{\rm line}\sim 8$\,keV that has been found in the jetted TDE Sw\,J1644+57, which has been interpreted as highly ionized iron K$\alpha$ emission blueshifted by $\sim0.15c$ \citep{Kara2016, Thomsen2019}. Instead, what is shown here indicate the possible existence of a relativistically broadened iron line (either redshifted or with a more distorted red wing). 

\paragraph{Modeling with Disk Reflection}
In this method, we fit the data using a combination of MCD and relativistic reflection from an accretion disk.

We utilize the self-consistent \texttt{relxill} model to describe the direct power-law component and the reflection part \citep{Garcia2014, Dauser2014}. 
In \texttt{relxill}, we fixed the outer disk radius ($R_{\rm out}$) at a fiducial value of $400\,R_{\rm g}$, since it has little effect on the X-ray spectrum. 
The redshift parameter in \texttt{relxill} was fixed at 0 since the host redshift was already included by the \texttt{zashift} model. 
To reduce the complexity of this model, we froze the reflection fraction $R_{\rm F}$ (ratio of the reflected to primary emission; \citealt{Dauser2016}) at 1. The inner and outer emissivity index $q$ were fixed at 3 throughout the accretion disk, making $R_{\rm break}$ obsolete. We assume the inner disk radius is at the innermost stable circular orbit (ISCO), i.e., $R_{\rm in}=R_{\rm ISCO}$.
Other parameters in \texttt{relxill} include the power-law index of the incident spectrum $\Gamma$, the cutoff energy of the power-law $E_{\rm cut}$, the black hole spin $a$, the inclination $i$, the ionization of the accretion disk $\xi$, the iron abundance of the accretion disk $A_{\rm Fe}$, and the normalization parameter $\rm Norm_{\rm rel}$. 
We first fit the data allowing $a$ to be free, finding that the fit is not sensitive to $a$. Therefore, we performed the fit with $a$ fixed at zero. 

\begin{figure}[htbp!]
    \centering
    \includegraphics[width=\columnwidth]{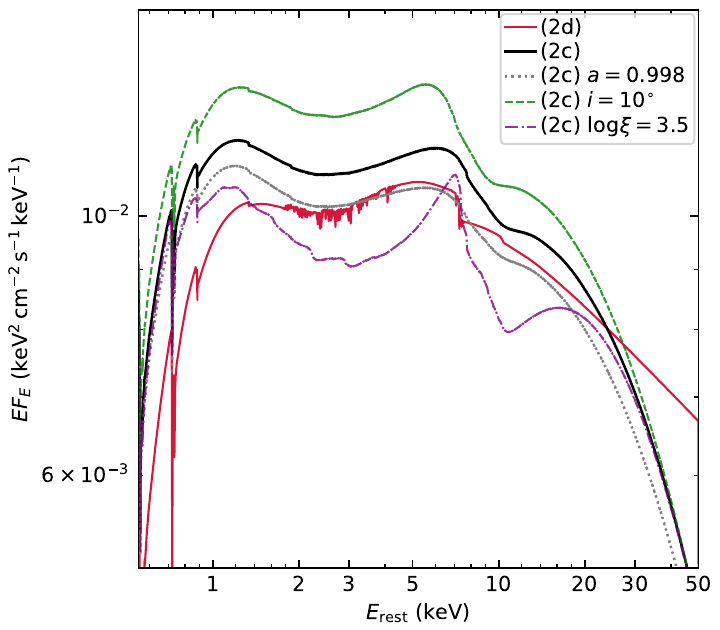}
    \caption{Best-fit incident model spectra of (2c) and (2d), as well as modifications of (2c) if one parameter is changed. }
    \label{fig:joint2_highres}
\end{figure}

The best-fit model, hereafter (2c), gives $\chi^2/dof =  306.65/296=1.04$ and is shown in the lower-left panel of Figure~\ref{fig:joint2_mods}. 
The best-fit model parameters are given in Table~\ref{tab:joint2_pars}. 
In Figure~\ref{fig:joint2_highres}, the solid black line shows the best-fit model; Modifications of the best-fit model are shown as dotted (if $a$ is changed from 0 to 0.998), dashed (if $i$ is changed from $43.4^{\circ}$ to $70^{\circ}$), and dash-dotted (if log$\xi$ is changed from 4.09 to 3.50) lines. The shape and width of the extremely broad iron emission are mainly determined by the high disk ionization state and the moderate inclination.
We note that the best-fit ionization of $\xi\sim 10^4\,{\rm erg\,cm\,s^{-1}}$ is greater than the typical values observed in Seyfert 1 AGN \citep{Walton2013, Ezhikode2020}.


\paragraph{Modeling with Absorbers}
In this method, we attempt to improve the fit by adding absorption features.
First, we added partial covering of a neutral absorber using the \texttt{pcfabs} model. In \texttt{pcfabs}, a fraction $f_{\rm cover}$ of the X-ray source is seen through a neutral absorber with hydrogen-equivalent column density $N_{\rm H}$, while the rest is assumed to be observed directly. The best-fit model gives $\chi^2 / dof = 386/297=1.30$. If we add a new free parameter (redshift of the neutral absorber) by replacing \texttt{pcfabs} with \texttt{zpcfabs}, $\chi^2/dof$ becomes $370/296=1.25$. However, both models leave 5--8\,keV flux excess in the residual.

Therefore, we next allow the absorber to be partially ionized by replacing \texttt{zpcfabs} with a photoionized absorber. This is also motivated by the fact that a good fit to the \chandra LETG observation conducted on 2021 November 29 was found with such a model by \citet{Miller2022_cxo}. This fit utilized the \texttt{zxipcf} model \citep{Reeves2008}, which is a grid of photoionization models computed by the \texttt{XSTAR} code \citep{Kallman2001}. 
However, \citet{Reynolds2012} noted that \texttt{zxipcf} only has a very coarse sampling in ionization space, 
and so in this work, we use an updated \texttt{XSTAR} grid that is suitable for use with AGNs (computed in \citealt{Walton2020}). 
This grid assumes an ionizing continuum of $\Gamma=2$ and a velocity broadening of $100\,{\rm km\,s^{-1}}$, and allows the ionization parameter, column density, absorber redshift, and both the oxygen and iron abundances to be varied as free parameters (although for simplicity we assume these abundances are solar). 
Fitting the data with the redshift of the absorber fixed at zero yields $\chi^2/dof=318.2/296=1.08$. If the redshift is allowed to be free, we have $\chi^2/dof = 317.7/295=1.08$. Since $\chi^2$ only reduces by 0.5 for 1 $dof$, the redshift parameter cannot be well constrained by our data. 
Therefore, we name the model with the absorber redshift fixed at zero as (2d), and show it in the lower-right panel of Figure~\ref{fig:joint2_mods}. 
The model parameters are given in Table~\ref{tab:joint2_pars}.
In  Figure~\ref{fig:joint2_highres}, the solid crimson line shows a high-resolution version of model (2d).

\paragraph{Model Comparison and Comments}
Between (2b) and (2c), we consider (2c) to be superior since (\textit{i}) its $\chi^2$ is smaller by 24 for only 1 $dof$ and (\textit{ii}) it adopts a physically motivated model instead of a mathematical function.

To compare (2c) and (2d), we use the Bayesian information criterion (BIC) to assess the goodness of fit. Here
\begin{align}
    {\rm BIC} &= k \cdot {\rm ln}(N) -2 {\rm ln}\mathcal{ L} \\
                &= k \cdot{\rm ln}(N) + \chi^2 + {\rm constant}
\end{align} 
where $k$ is the number of free parameters, $N$ is the number of spectral bins, and $\mathcal{L}$ is the maximum of the likelihood function. 
Models with lower BIC values are favored. 
According to \citet{Raftery1995}, a BIC difference between 2 and 6 is positive, a difference between 6 and 10 is strong, and a difference greater than 10 is very strong.
Since ${\rm BIC(2c)} - {\rm BIC(2d)} =-5.9$, model (2c) is slightly favored over (2d). 
The energy range over which model (2c) performs better than (2d) is $\sim 8$--12\,keV. 
This is because absorption by ionized iron adds a relatively sharp flux decrease at $\sim7$\,keV, while the blue wing of the iron emission in \texttt{relxill} is smoother (see the lower-right panel of Figure~\ref{fig:joint2_mods}). 

We note that the residual below 0.7\,keV is strong in all model fits, and is likely caused by underestimated \nicer calibration uncertainties at the lowest energies.

\subsubsection{\xmm Analysis} \label{subsubsec:XMMspec}

\begin{deluxetable}{llcc}[htbp!]
\tablecaption{Modeling of two \xmm observations. 
\label{tab:XMM_fit} }
\tablehead{
\colhead{Component} 
&  \colhead{Parameter} 
& \colhead{XMM\,E1}
& \colhead{XMM\,E2} 
}
\startdata
\texttt{ztbabs} & $N_{\rm H}$ ($10^{20}\,{\rm cm^{-2}}$) 
& $1.09^{+0.99}_{-0.45}$ 
& $<1.22$ 
\\
\texttt{diskbb} & $T_{\rm in}$ (eV) 
& $68^{+1}_{-4}$  
& $125\pm 8$ 
\\
& $R_{\rm in}^\ast$ ($10^4\,{\rm km}$) 
& $511^{+144}_{-75}$    
& $39^{+10}_{-6}$   
\\
\texttt{simpl} & $\Gamma$
& $>4.57$ \tablenotemark{$\dagger$}
& $2.92\pm0.15$ 
\\
& $f_{\rm sc}$
& $0.13^{+0.03}_{-0.01}$  
& $0.16\pm0.03$ 
\\
--- & $\chi^2/dof$
& $70.26/52=1.35$
& $97.49/82=1.19$
\\
\enddata
\tablenotetext{\dagger}{Upper limit of $\Gamma$ is at 5.}
\end{deluxetable}

\begin{figure}[htbp!]
    \centering
    \includegraphics[width=\columnwidth]{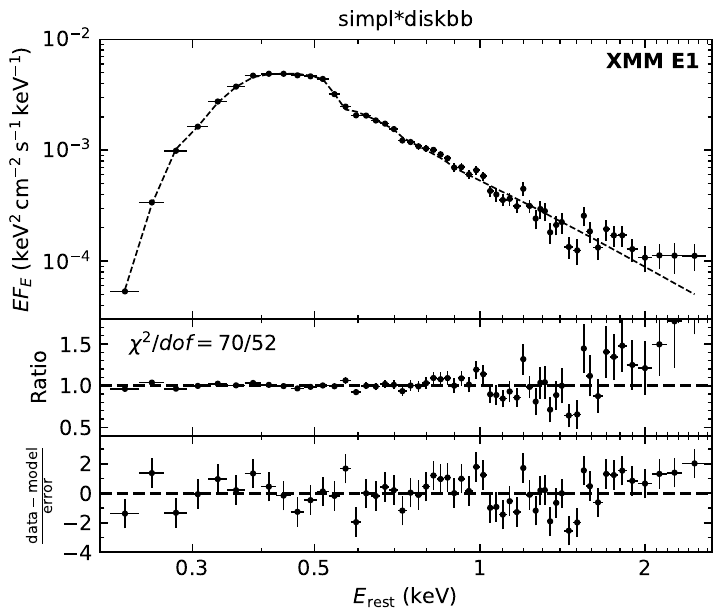}
    \includegraphics[width=\columnwidth]{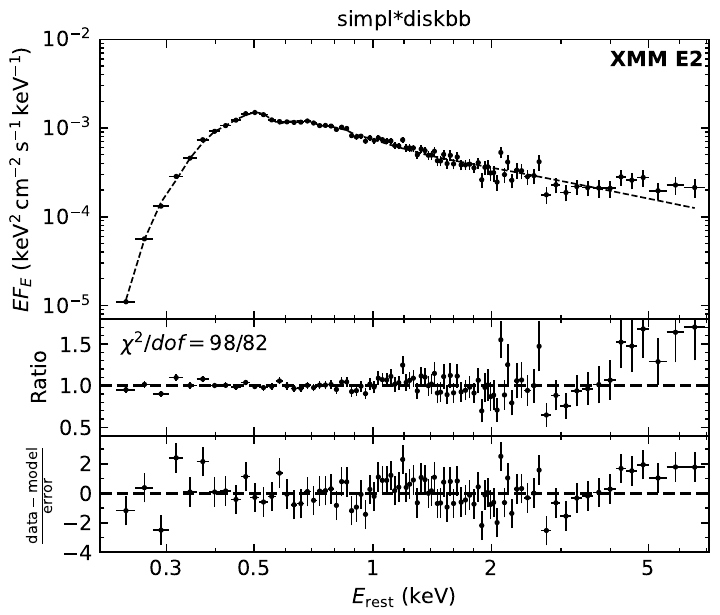}
    \caption{The \xmm spectra. 
    The data have been rebinned for visual clarity. 
    The dashed lines show the best-fit models.
    See Table~\ref{tab:XMM_fit} for the best-fit parameters.\label{fig:xmm_spec}}
\end{figure}

We chose energy ranges where the source spectrum dominates over the background. For XMM\,E1 this is 0.2--2.6\,keV, while for XMM\,E2 this is 0.2--7.0\,keV. All data were fitted using $\chi^2$-statistics. Following \S\ref{subsubsec:joint1} and \S\ref{subsubsec:joint2}, all models described below have been multiplied by \texttt{tbabs*ztbabs*zashift} to include Galactic absorption, host absorption, and host redshift. 

Although the XMM\,E1 spectrum is very soft, a single MCD results in a poor fit and leaves a large residual above 1\,keV, suggesting the existence of a non-thermal component. A continuum model of \texttt{simpl*diskbb} gives a much better fit with $\chi^2_{\rm r} = 1.35$. The best-fit model is shown in the upper panel of Figure~\ref{fig:xmm_spec}.
The XMM\,E2 spectrum is much harder than that from XMM\,E1. 
Fitting with \texttt{simpl*diskbb} gives a good fit with $\chi^2_{\rm r} = 1.19$ (see Figure~\ref{fig:xmm_spec}, lower panel). 

We note that although the $\chi^2_{\rm r}$ of our best-fit \xmm models are acceptable, there seems to be some systematic residuals. 
For example, eight consecutive bins of positive residuals are seen in the 1.7--2.6\,keV XMM\,E1 data. 
A possible explanation is that there exist spectral features created by absorbing materials in the TDE system, such as the blueshifted absorption lines reported in the TDE ASASSN-14li \citep{Miller2015_disk_wind, Kara2018}.
Seven consecutive bins of positive residuals are seen in the 4.0--7.0\,keV XMM\,E2 data. This might indicate the existence of  disk reflection features, such as an iron emission line. We note that 2--4\,days after our second \xmm epoch, \xmm/EPIC observations obtained under another program also reveals the existence of interesting features in the iron K band  \citep{Miller2022_xmm}.
More detailed modeling of the \xmm spectra is beyond the scope of this paper, and is encouraged in future work. 

\subsubsection{\srge Analysis} \label{subsubsec:SRGspec}

\begin{figure}[htbp!]
    \centering
    \includegraphics[width=\columnwidth]{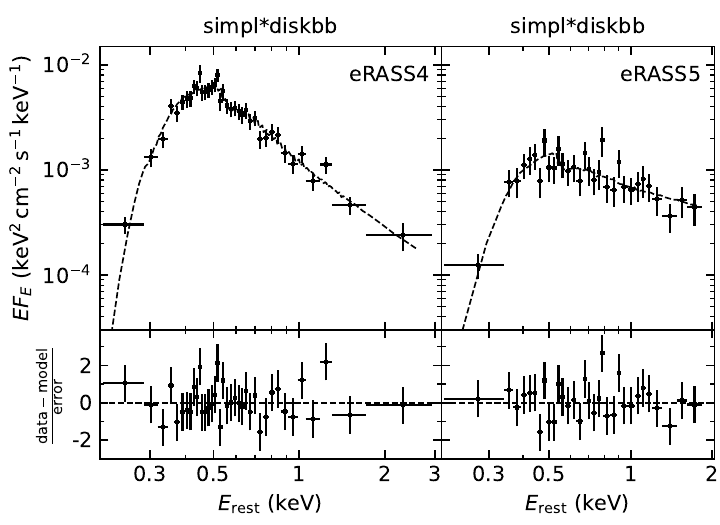}
    \caption{\srg/eROSITA spectra of \target. \label{fig:srg_spec}}
\end{figure}

\begin{deluxetable}{llcc}
\tablecaption{Modeling of two \srge observations. 
\label{tab:srg_spec} }
\tablehead{
\colhead{Component} 
&  \colhead{Parameter} 
& \colhead{eRASS4}
& \colhead{eRASS5} 
}
\startdata
\texttt{ztbabs} & $N_{\rm H}$ ($10^{20}\,{\rm cm^{-2}}$) 
& $0.21_{-0.20}^{+2.19}$ 
& $<3.41$
\\
\texttt{diskbb} & $T_{\rm in}$ (eV) 
& $89^{+7}_{-13}$  
& $96_{-22}^{+32}$ 
\\
& $R_{\rm in}^\ast$ ($10^4\,{\rm km}$) 
& $210^{+179}_{-38}$    
& $73^{+243}_{-21}$   
\\
\texttt{simpl} & $\Gamma$
& $4.15^{+0.82}_{-0.73}$ 
& 2.92 (frozen) 
\\
& $f_{\rm sc}$
& $0.14^{+0.12}_{-0.08}$  
& $0.21^{+0.06}_{-0.09}$  
\\
--- & $cstat/dof$
& $126.43/140$
& $76.45/85$
\\
\enddata
\end{deluxetable}

We chose energy ranges where the source spectrum dominates over the background. For eRASS4 this range is 0.2--3\,keV, while for eRASS5 this range is 0.2--2\,keV. All data were fitted with the $C$-statistic \citep{Cash1979}. 

Following \S\ref{subsubsec:joint1} and \S\ref{subsubsec:XMMspec}, we fitted the \srge spectra with \texttt{tbabs*ztbabs*zashift*
simpl*diskbb}. For the eRASS5 spectrum, since the source is only
above background at 0.2--2\, keV, the powerlaw index cannot be constrained from the \srge spectrum alone. Therefore, we fixed $\Gamma$ at the best-fit value of the XMM\,E2 spectrum (Table~\ref{tab:XMM_fit}, $\Gamma = 2.92$), and allowed other parameters to be free. This choice is based on the fact that the XMM\,E2 and eRASS5 observations appear to show the same properties on the light curve and hardness evolution diagrams (Figure~\ref{fig:xray_lc}).

The fitting results are shown in Figure~\ref{fig:srg_spec}. 
The best-fit parameters are shown in Table~\ref{tab:srg_spec}.

\subsubsection{XRT Analysis} \label{subsubsec:XRTspec}
\begin{figure}[htbp!]
    \centering
    \includegraphics[width=\columnwidth]{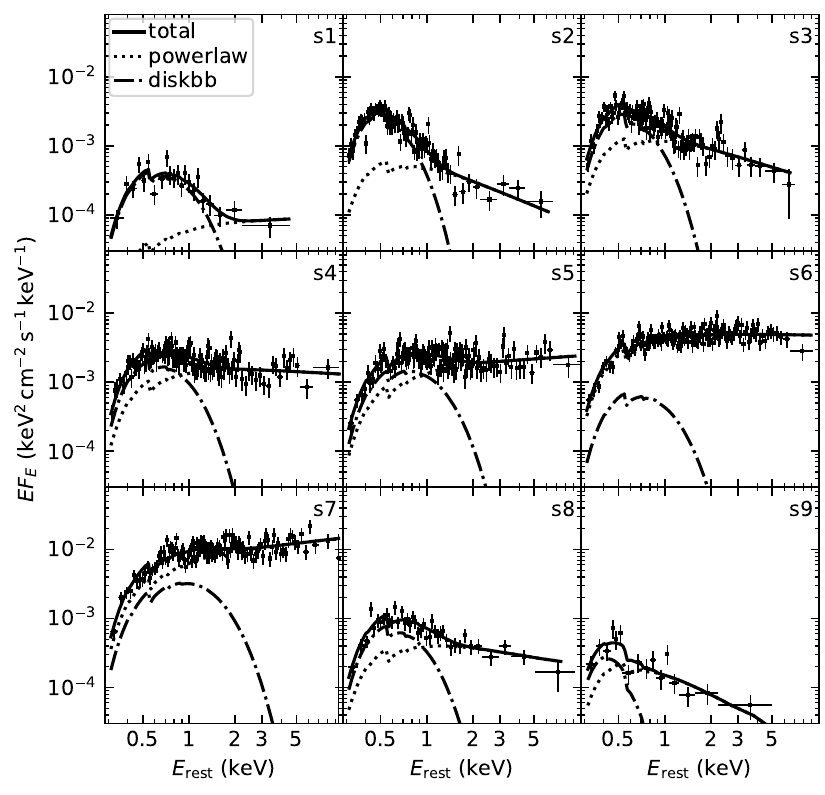}
    \caption{XRT time-averaged spectra of \target. 
    See the lower panel of Figure~\ref{fig:xray_lc} for the time span of each spectrum.
    \label{fig:xrt_spec}}
\end{figure}

The temporal coverage of each time-averaged XRT spectrum (generated in \S\ref{subsubsec:XRTobs}) is shown as `s1', `s2', ..., `s9' in the lower panel of Figure~\ref{fig:xray_lc}. 
We fitted the 0.3--10\,keV spectra using a simple model of \texttt{tbabs*zashift*(diskbb+powerlaw)}. 
We did not include the \texttt{ztbabs} component, as host galaxy absorption was found to be negligible or much smaller than the Galactic absorption in all previous spectral analyses (see Tables~\ref{tab:joint1_pars}, \ref{tab:joint2_pars}, \ref{tab:XMM_fit}, and \ref{tab:srg}). 
The adopted continuum model does not give realistic model parameters. For example, the disk radii will be underestimated when the source spectrum is hard (see a detailed discussion in \citealt{Steiner2009}). 
The main goal of this fitting is to compute the multiplicative factor to convert the 0.3--10\,keV XRT net count rate to X-ray fluxes, including 
\ad{(i) the observed 0.3--10\,keV flux $f_{\rm X}$(0.3--10\,keV), 
(ii) the Galactic absorption corrected 0.3--10\,keV flux $f_{\rm X, 0}$ (0.3--10\,keV), 
(iii) the Galactic absorption corrected 0.5--10\,keV flux $f_{\rm X, 0}$ (0.5--10\,keV),
(iv) the Galactic absorption corrected flux density at the rest-frame energies of 0.5\,keV and 2\,keV \ad{(i.e., $f_\nu$(0.5\,keV) and $f_\nu$(2\,keV))}.}  
All data were fitted using $C$-statistics. 

The best-fit models are shown in Figure~\ref{fig:xrt_spec}. 
\ad{Scaling factors to convert 0.3--10\,keV net count rate to X-ray fluxes can be computed using values provided in Table~\ref{tab:xrtfit} (Appendix~\ref{subsec:modelfit}).}
The observed isotropic equivalent 0.3--10\,keV X-ray luminosity, $L_{\rm X}$, is shown in the upper panel of Figure~\ref{fig:multi_evol}. Note that for the initial four XRT non-detections, we assume a spectral shape similar to `s1'.

\subsubsection{\nicer Analysis} \label{subsubsec:NICERspec}
We started with the obsID-binned \nicer spectra generated in \S\ref{subsec:NICERobs}. 
We only performed spectral fitting on obsIDs with more than 500 total net counts in 0.3--4\,keV. 
Following \S\ref{subsubsec:XRTspec}, we fitted a \texttt{tbabs*zashift*(diskbb+powerlaw)} model to the 0.3--4\,keV spectra and inferred $f_{\rm X}$ from the best-fit models. 
All data were fitted using $\chi^2$-statistics. 
The best-fit models provided a $\chi^2_{\rm r}$ close to 1 in most cases. 
The $L_{\rm X}$ evolution inferred from \nicer spectral fitting is also shown in the upper panel of Figure~\ref{fig:multi_evol}.

\subsection{Spectral Indices $\alpha_{\rm OX}$ and $\alpha_{\rm OSX}$} \label{subsec:alpha_OX}

\begin{figure}[htbp!]
    \centering
    \includegraphics[width=\columnwidth]{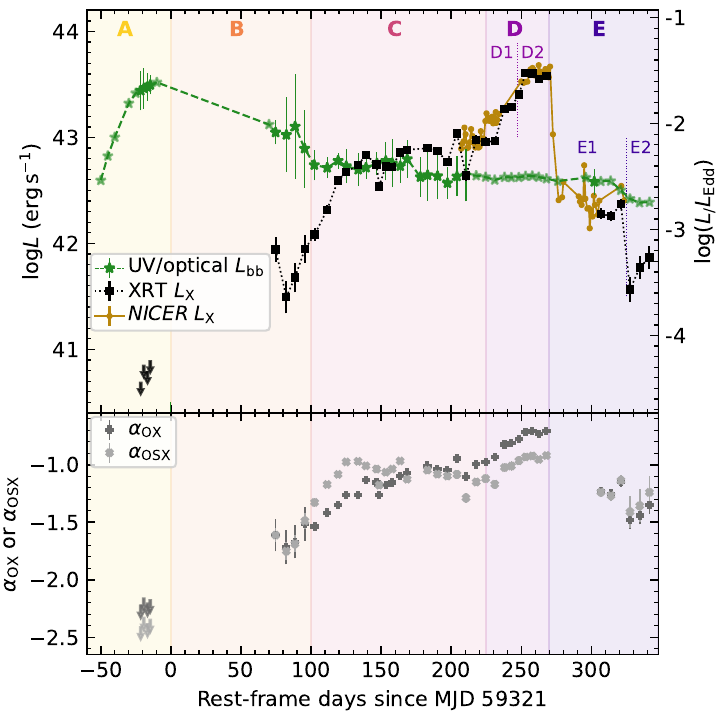}
    \caption{\textit{Upper}: blackbody luminosity of the UV/optical emission ($L_{\rm bb}$; \S\ref{subsec:uvopt_bbfit}) compared with the observed isotropic equivalent 0.3--10\,keV X-ray luminosity ($L_{\rm X}$) from XRT (\S\ref{subsubsec:XRTspec}) and \nicer (\S\ref{subsubsec:NICERspec}). 
    \textit{Lower}: the 2500\,\AA\ to X-ray spectral slope measured by \swift observations (Eq.~\ref{eq:alpha_OX},~\ref{eq:alpha_OSX}). 
    \label{fig:multi_evol}}
\end{figure}

\begin{figure}[htbp!]
    \centering
    \includegraphics[width=\columnwidth]{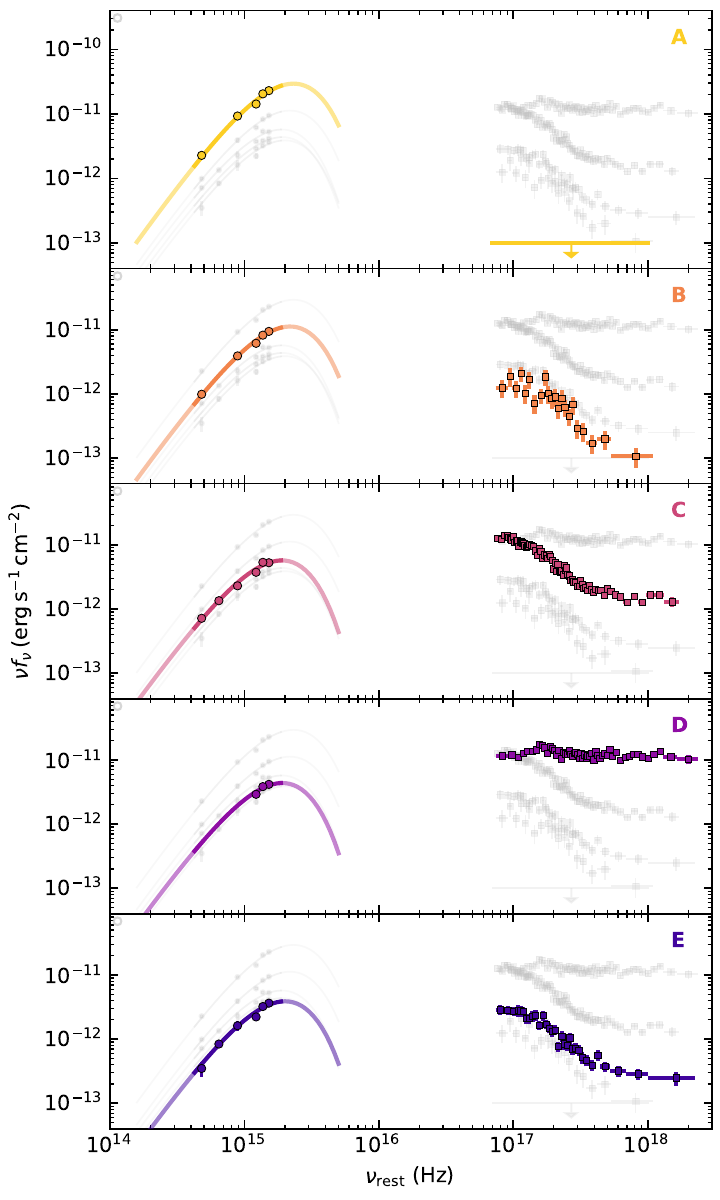}
    \caption{Typical SEDs of \target in five phases. The data has been corrected for extinction (in UV/optical) and column density absorption (in the X-ray). The solid lines are the blackbody fits to UV/optical data. 
    \label{fig:sed_evol}}
\end{figure}

To assist comparison with TDEs from the literature, we computed the UV to X-ray spectral index $\alpha_{\rm OX}$ \citep{Tananbaum1979, Ruan2019, Wevers2021} and $\alpha_{\rm OSX}$ \citep{Gezari2021}, which are commonly used in AGN and TDE literature to characterize the ratio of UV to X-ray fluxes\footnote{Note that some papers use these indices with a minus sign in front of our definitions.}. 
Here 
\begin{align}
    \alpha_{\rm OX} & \equiv \frac{{\rm log}[L_\nu(2500\,{\rm \AA}) / L_\nu(2\,{\rm keV})]}{{\rm log}[\nu(2500\,{\rm \AA}) / \nu(2\,{\rm keV})]} \label{eq:alpha_OX},\\
    \alpha_{\rm OSX} & \equiv \frac{{\rm log}[L_\nu(2500\,{\rm \AA}) / L_\nu(0.5\,{\rm keV})]}{{\rm log}[\nu(2500\,{\rm \AA}) / \nu(0.5\,{\rm keV})]} \label{eq:alpha_OSX},
\end{align}
where $L_\nu$ is the luminosity at a certain frequency (corrected for $N_{\rm H}$ and $E_{B-V, {\rm MW}}$).
We use the \swift $uvw1$ host-subtracted luminosities (rest-frame effective wavelength at $2459\,{\rm \AA}$ for $T_{\rm eff}=3\times10^4\,{\rm K}$) as a proxy for $L_\nu (2500\,{\rm \AA})$. 
We measure $f_\nu (0.5\,{\rm keV})$ and $f_\nu (2\,{\rm keV})$ by converting the XRT net count rates to flux densities using the scaling factors derived in \S\ref{subsubsec:XRTspec}. 
We note that $f_\nu (2\,{\rm keV})$ mainly traces the evolution of the non-thermal X-ray component, while $f_\nu (0.5\,{\rm keV})$ traces both the thermal and non-thermal components. The results are shown in the lower panel of Figure~\ref{fig:multi_evol}.

Based on Figure~\ref{fig:multi_evol}, we divide the evolution of \target into five phases.
In phase A ($\delta t \lesssim 0$\,days), the UV/optical luminosity brightens, while X-rays are not detected ($< 10^{40.9}\,{\rm erg\,s^{-1}}$). 
In phase B ($0\lesssim \delta t \lesssim  100$\,days), the UV/optical luminosity declines, and X-rays emerge. 
Entering into phase C ($100 \lesssim \delta t \lesssim 225$\,days), the X-ray spectrum gradually hardens, while the UV/optical luminosity stays relatively flat. 
In phase D ($225 \lesssim \delta t \lesssim 270$\,days), the X-ray further brightens two times (indicated by D1 and D2), and the UV/optical plateau persists. 
In phase E, the X-ray luminosity drops two times (indicated by E1 and E2), while the UV/optical luminosity only slightly declines. 
Interestingly, the dramatic X-ray evolution in phase D+E does not have much effect on the UV/optical luminosity. Typical SEDs in each phase are shown in Figure~\ref{fig:sed_evol}.

\subsection{Bolometric Luminosity $L_{\rm bol}$}

To calculate the bolometric luminosity $L_{\rm bol}$ at the epochs of the \swift observations, we assume that the bulk of the radiation is between 10000\,\AA\ and 10\,keV.
We estimate that when the X-ray spectrum is the hardest (i.e., model 2c), the 0.3--10\,keV flux still constitutes 72\% of the 0.3--100\,keV flux. 
Therefore, this assumption at most underestimates log$L_{\rm bol}$ by 0.14 dex. 

\begin{figure}[htbp!]
    \centering
    \includegraphics[width = \columnwidth]{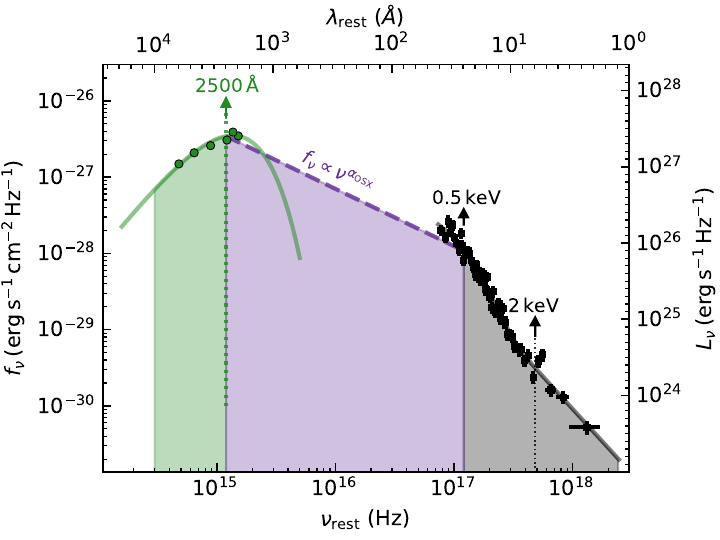}
    \caption{A snapshot SED of \target at $\delta t\approx 147$\,days. The data has been corrected for extinction (in UV/optical) and Galactic absorption (in the X-ray). 
    The solid lines are the blackbody fits to UV/optical data (\S\ref{subsec:uvopt_bbfit}) and the XRT `s3' spectrum best-fit model (\S\ref{subsubsec:XRTspec}).
    The shaded region shows that the $L_{\rm bol}$ is calculated in three energy ranges (see text).\label{fig:demo_calc_Lbol}}
\end{figure}

We compute the 10000\,\AA\ to 10\,keV luminosity by adding the luminosities in three energy ranges (see a demonstration in Figure~\ref{fig:demo_calc_Lbol}). 

From 10000\,\AA\ to 2500\,\AA, we integrate below the blackbody model fitted to the UV/optical photometry (\S\ref{subsec:uvopt_bbfit}).

From 2500\,\AA\ to 0.5\,keV, we assume that the TDE spectrum is continuous and can be approximated by a power-law of $f_\nu \propto \nu^{\alpha_{\rm OSX}}$. 
Hence, the luminosity is 
\begin{align}
    L & = \int_{\nu_1}^{\nu_2} L_\nu d \nu 
    \approx \int_{\nu_1}^{\nu_2} \frac{L_\nu(\nu_1)}{\nu_1^{\alpha_{\rm OSX}}} \nu^{\alpha_{\rm OSX}}  d \nu \\
    & = \frac{L_\nu (\nu_1)}{\nu_1^{\alpha_{\rm OSX}}} \times
    \begin{cases}
    \displaystyle
      \frac{\nu_2^{\alpha_{\rm OSX}+1} - \nu_1^{\alpha_{\rm OSX}+1}}{\alpha_{\rm OSX}+1} & \text{if $\alpha_{\rm OSX} \neq -1$}\\
      {\rm ln}(\nu_2 / \nu_1) & \text{if $\alpha_{\rm OSX} = -1$}
    \end{cases}   
\end{align}
where $\nu_1 = 10^{15.08}\,{\rm Hz}$, $\nu_2 = 10^{17.08}\,{\rm Hz}$. In this range, we assume that the uncertainty of $L$ is $0.3L$. 

From 0.5\,keV to 10\,keV, we calculate the luminosity by converting the 0.3--10\,keV XRT net count rate to Galactic absorption corrected 0.5--10\,keV luminosity using the scaling factors derived in \S\ref{subsubsec:XRTspec}.

Note that for the first four \swift epochs, since X-rays were not detected, we use the UV/optical blackbody luminosity $L_{\rm bb}$ as an approximation of $L_{\rm bol}$. 

\begin{figure}[htbp!]
    \centering
    \includegraphics[width = \columnwidth]{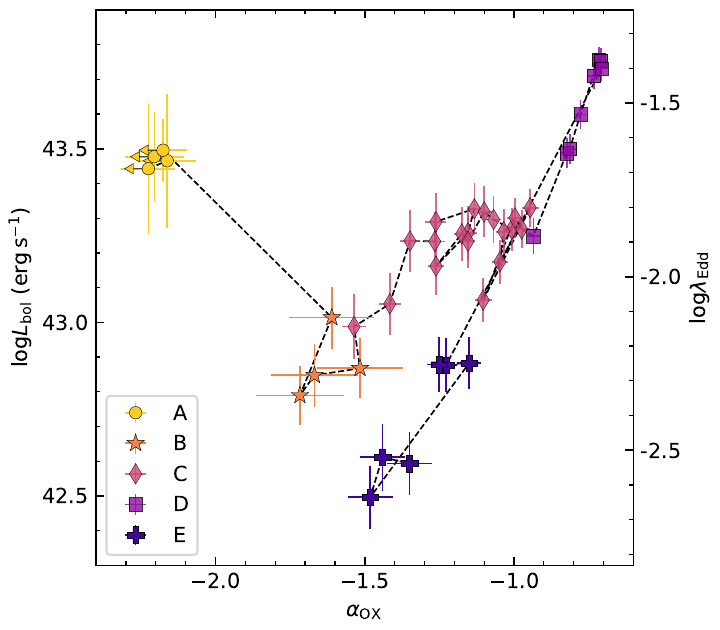}
    \caption{The bolometric luminosity $L_{\rm bol}$ as a function of $\alpha_{\rm OX}$. \ad{$L_{\rm bol}$ is converted to $\lambda_{\rm Edd}$ assuming ${\rm log}(M_{\rm BH}/M_\odot )=7.03$ (\S\ref{subsec:esi}).}
    Note that the uncertainty of $\lambda_{\rm Edd}$ should be greater than the uncertainty of $L_{\rm bol}$ by 0.44\,dex (i.e., the uncertainty of $M_{\rm BH}$), which is not included in the figure.
    \label{fig:Lbol_alpha}}
\end{figure}

The evolution of log$L_{\rm bol}$ as a function of $\alpha_{\rm OX}$ is shown in Figure~\ref{fig:Lbol_alpha}. The data points are color coded by their phases (from A to E, see Figure~\ref{fig:multi_evol}). The right $y$-axis converts $L_{\rm bol}$ to the Eddington ratio $\lambda_{\rm Edd}\equiv L_{\rm bol}/L_{\rm Edd}$. For pure hydrogen, a $M_{\rm BH}$ of $10^{7.03}\,M_\odot$ (\S\ref{subsec:esi}) implies an Eddington luminosity of $L_{\rm Edd} \approx 10^{45.13}\,{\rm erg\,s^{-1}}$. We further discuss this figure in \S\ref{subsec:disk_corona}.
\ad{The maximum luminosity was reached at $\delta t = 253$\,days, with $L_{\rm bol}=(7.94\pm 0.66)\times 10^{43}\,{\rm erg\,s^{-1}}$ and $\lambda_{\rm Edd} = 6.0^{+10.4}_{-3.8}$\%.} 
As a cautionary note, the relatively low value of $\lambda_{\rm Edd}$ (\ad{$<16$\%}) does not necessarily imply that the accretion is in the sub-Eddington regime, as the TDE broadband SED may peak in the extreme-UV (EUV) band \citep{Dai2018, Mummery2020}.
 
\section{Discussion} \label{sec:discussion}
Hereafter we define $M_7 \equiv M_{\rm BH} / (10^7\,M_\odot)$, $\dot m \equiv \dot M_{\rm acc} / \dot M_{\rm Edd}$, $\dot M_{\rm Edd} \equiv L_{\rm Edd} / (\eta c^2)$, $\eta_{-1}\equiv \eta/10^{-1}$, where $\dot M_{\rm acc}$ is the mass accretion rate and $\eta$ is the accretion radiative efficiency. 
With $M_7 \approx 1$, the gravitational radius is $R_{\rm g} =G M_{\rm BH}/c^2 \approx 10^{12.20}\,{\rm cm}$. For a solar type star, the tidal radius is $R_{\rm T} = 10^{13.19}\,{\rm cm}\approx 10R_{\rm g}$, within which the tidal force exceeds the star's self-gravity \citep{Rees1988}.

\subsection{Origin of the Soft X-ray Emission} \label{subsec:origin_softXray}
\begin{figure}[htbp!]
    \centering
    \includegraphics[width=\columnwidth]{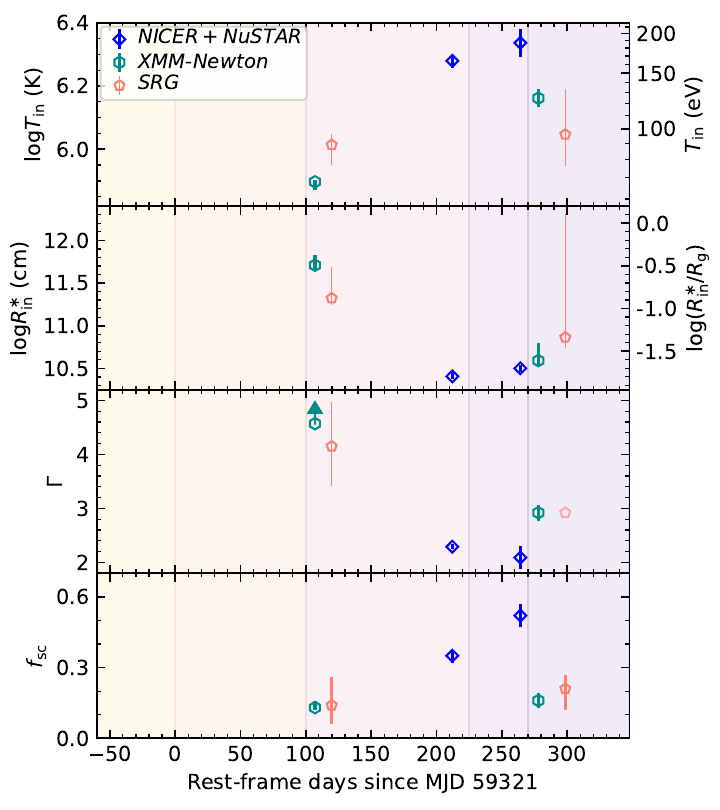}
    \caption{Evolution of best-fit X-ray spectral parameters, including log$T_{\rm in}$ and log$R_{\rm in}^{\ast}$ in the \texttt{diskbb} component (top two panels), $\Gamma$ and $f_{\rm sc}$ in the \texttt{simpl} component (third and fourth panels). 
    Note that the uncertainty of log($R_{\rm in}^{\ast}/R_{\rm g}$) is greater than the uncertainty of $R_{\rm in}^{\ast}$ by 0.44\,dex (i.e., the uncertainty of $M_{\rm BH}$; \S\ref{subsec:esi}), which is not included in the figure.
    Data are from model (1a) in Table~\ref{tab:joint1_pars}, model (2a) in Table~\ref{tab:joint2_pars}, Table~\ref{tab:XMM_fit}, and Table~\ref{tab:srg_spec}. 
    For parameters in model (2a), we assume an uncertainty of 10\%.
    Fixed values are shown as semi-transparent symbols. 
    Background colors follow the scheme shown in Figure~\ref{fig:multi_evol}.
    \label{fig:xpar_evol}}
\end{figure}

The soft X-ray emission of many TDEs has been attributed to the inner regions of an accretion disk \citep{Saxton2020}. 
Assuming $R_{\rm in}\approx 6 R_{\rm g} \approx 10^{13}\,{\rm cm}$, the maximum effective temperature of an optically thick, geometrically thin accretion disk is $T_{\rm eff}\approx 20 (\frac{\dot m}{M_7\eta_{-1}})^{1/4}$\,eV \citep{Shakura1973}.
With a maximum black hole spin of $a \rightarrow 1$, $R_{\rm in} \rightarrow R_{\rm g}$, and $T_{\rm eff}\approx 78 (\frac{\dot m}{M_7\eta_{-1}})^{1/4}$\,eV. 
The top panel of Figure~\ref{fig:xpar_evol} shows that in phase D, when the X-ray spectrum is the hardest, the measured $T_{\rm in}$ is $\sim 2.5$ times greater than the maximum allowed $T_{\rm eff}$. 
This high color temperature causes the inferred disk radius $R_{\rm d,in}=R_{\rm in}^{\ast}/\sqrt{{\rm cos} i}$ to be much less than $R_{\rm g}$ throughout the evolution. 
The projection factor $\sqrt{\cos i}$ should not be much less than unity, since for a nearly edge-on viewing angle, the X-rays from the inner disk will be obscured by the gas at larger radii. 
The relativistic disk reflection model (2c) also suggests $\sqrt{\cos i}=0.85^{+0.06}_{-0.07}$.
We note that disk radii much less than $R_{\rm g}$ have also been inferred in a few other X-ray bright TDEs (see, e.g., Fig.~8 of \citealt{Gezari2021}).

This discrepancy may be due to Compton scattering \citep{Shimura1995}, which makes the measured temperature greater than the effective inner disk temperature by a factor of $f_c$ \citep{Davis2019}, i.e., $T_{\rm in}=f_c T_{\rm eff}$. The physical reason is that, as the X-ray photons propagate in the vertical direction away from the disk mid-plane, the color temperature is determined by the thermalization depth (corresponding to the last absorption), which could be located at a high scattering optical depth $\tau\gg 1$ --- this causes $T_{\rm in}$ to be higher than $T_{\rm eff}$ by a factor of $\sim \tau^{1/4}$. As a result of ongoing fallback, the vertical structure of the TDE disk \citep[see][]{Bonnerot2021_disk_formation} is likely substantially different from the standard thin disk as studied by \citet[][who concluded $f_{\rm c}\lesssim 2$]{Davis2019}, so the color correction factor may be different. More detailed radiative transport calculations in the TDE context are needed to provide a reliable $f_{\rm c}$ based on first principles. 

Another possible reason for the seemingly small disk radii is that a scattering dominated, Compton-thick gas layer can suppress the X-ray flux without causing any significant change to the spectral shape. 
For a spatially uniform layer, the transmitted flux is exponentially suppressed for a large scattering optical depth $\gg 1$. A more likely configuration is that the layer is like an obscuring wall with small holes where a fraction of the source X-rays can get through, and the rest of the area contributes negligibly to the observed flux. 
In this scenario, the inferred disk radius is reduced by a factor of the square root of the transmitted over emitted fluxes.

\subsection{Implications of the Hard X-ray Emission} \label{subsec:hardXray}

Hard X-rays can be generated by Compton up-scattering of soft X-rays from the accretion disk by the hot electrons in the (magnetically dominated) coronal regions above the disk, as is the case in AGNs and XRBs. The physical situation in TDEs is more complicated than in AGNs in that the hard X-rays must make their way out of the complex hydrodynamic structures. An X-ray photon undergoes $\sim\tau^2$ electron scatterings as it propagates through a gas slab of Thomson optical depth $\tau$. In each scattering, the photon loses a fraction $E_\gamma /m_e c^2$ of its energy (where $E_\gamma$ is the photon energy) as a result of Compton recoil, and hence the cumulative fractional energy loss is $\sim \tau^2 E_\gamma /m_e c^2$. This means that photons above an energy threshold of $\sim 1\mr{\,keV} (\tau/20)^{-2}$ will be Compton down-scattered by the gas.

Our \nustar observations clearly detected hard X-ray photons up to $30\rm\, keV$, which requires that the optical depth along the pathways of these photons from the inner disk ($\gtrsim R_{\rm g} \sim 10^{12.2}\rm\,cm$) to the observer is less than about 4. On the other hand, the UV/optical emission indicates that the reprocessing layer is optically thick up to a radius of the order of $R_{\rm bb}\sim10^{14}\rm\, cm$. Therefore, our observations favor a highly non-spherical system --- there are viewing angles that have very large optical depths such that most X-ray photons are absorbed (and reprocessed into the UV/optical bands), and there are other viewing angles with scattering optical depth $\tau \lesssim 4$ so that hard X-ray photons can escape. 

\subsection{X-ray Spectral Evolution}
\subsubsection{Soft to Hard Transition: Corona Formation} \label{subsubsec:soft-to-hard}
The top two panels of Figure~\ref{fig:xpar_evol} show that, during the soft $\rightarrow$ hard transition, AT2021ehb's inferred inner disk radius ``moves'' inward. We find that the main cause of this behavior is that the inner disk temperature increases with time as the spectrum hardens. The gradual hardening is consistent with a picture where it takes $\sim10^2$\,days to build up the magnetically dominated hot corona region. It is possible that the initially weak magnetic fields in the bound debris are amplified by differential rotation of the disk and the magnetorotational instability \citep{Balbus1991, Miller2000}.

\subsubsection{Hard to Soft Transition: \\ Thermal-viscous Instability?} \label{subsubsec:thick-thin-disk}

The rapid X-ray flux drop (D$\rightarrow $E) is likely due to a state transition in the innermost regions of the accretion disk. Under the standard $\alpha$-viscosity prescription where the viscous stress is proportional to the total (radiation $+$ gas) pressure \citep{Shakura1973}, the disk undergoes a thermal-viscous instability as the accretion rate drops from super- to sub-Eddington regimes \citep{lightman74_instability, shakura76_instability}. This instability causes the disk material to suddenly transition from radiation pressure-dominated to gas pressure-dominated state on a sound-crossing timescale, and the consequence is that the disk becomes much thinner and hence the accretion rate drops. \cite{shen14_disk_evolution} considered the thermal-viscous instability in the TDE context but concluded that the instability should occur within a few months since the disruption and the accretion rate drops by several orders of magnitude --- these, taken at face value, are inconsistent with our observations. More detailed work on the disk evolution is needed to draw a firm conclusion. Here, we provide two arguments for the disk state transition explanation.

First, TDEs with relativistic jets \citep[e.g.,][]{Bloom2011, Burrows2011, Cenko2012, Pasham2015} also show a sharp drop in X-ray luminosity 200 to 300 days (in the rest frame) after the discovery and that has been interpreted as the thick-to-thin transition of the inner disk \citep{tchekhovskoy14_mad_jet}. Second, from the mass fallback rate $\dot{M}_{\rm fb}\simeq M_*/3P_{\rm min} (t/P_{\rm min})^{-5/3}$ ($M_*$ being the stellar mass and $P_{\rm min}$ being the minimum period of the fallback material), one can estimate the time $t_{\rm Edd}$ at which the fallback rate drops below the Eddington accretion rate of $\sim 10L_{\rm Edd}/c^2$, and the result is \citep{Lu18}
\begin{equation}
    t_{\rm Edd}\simeq 309 \mr{\,d}\, M_{\rm h,7}^{-2/5} \left( \frac{M_\ast}{M_\odot} \right)^{\ad{ (1+3q)/5}},
\end{equation}
where we have taken the \ad{normal mass-radius relation of main sequence stars $R_\ast \simeq R_\odot (M_\ast/\msun)^q$ ($q=0.8$ below one solar mass stars and $q=0.57$ above one solar mass stars)}. 
We expect the inner disk to collapse into a thin state on the timescale of $t_{\rm Edd}$, under the condition that an order-unity fraction of the fallback rate directly reaches near the innermost regions of the disk. We note that the condition is satisfied for a $\approx 10^7M_\odot$ MBH since the tidal radius is only $\approx 10R_{\rm g}$ for a solar-type star.

If the rapid X-ray flux drop ($\rm D\rightarrow E$) is indeed caused by a disk state transition, then after the transition, the disk mass will accumulate over time due to ongoing fallback. This causes the accretion rate to increase, and eventually the disk briefly goes back to a thick state (with a very short viscous time) followed by another transition to the thin state. This is qualitatively consistent with the second rapid X-ray flux decline at $\delta t \approx 325\,$days ($\rm E1\rightarrow E2$ in Figure~\ref{fig:multi_evol}).

\subsection{Unusual UV/optical Behavior} 

\subsubsection{Featureless Optical Spectrum} \label{subsubsec:whynoline}

As shown in \S\ref{subsec:opt_spec_analyze}, \target's optical spectroscopic properties are dissimilar to the majority of previously known TDEs (i.e., H-rich, He-rich, N-rich, Fe-rich; \citealt{Leloudas2019, vanVelzen2021, Wevers2019_18fyk}). It is most similar to a few recently reported TDEs with blue and featureless spectra \citep{Brightman2021, Hammerstein2022}. \citet{Hammerstein2022} found that compared with TDEs that develop broad emission lines, the UV/optical emission of four featureless events have larger peak $L_{\rm bb}$, peak $T_{\rm bb}$, and peak $R_{\rm bb}$. 

\begin{figure}[htbp!]
    \centering
    \includegraphics[width=\columnwidth]{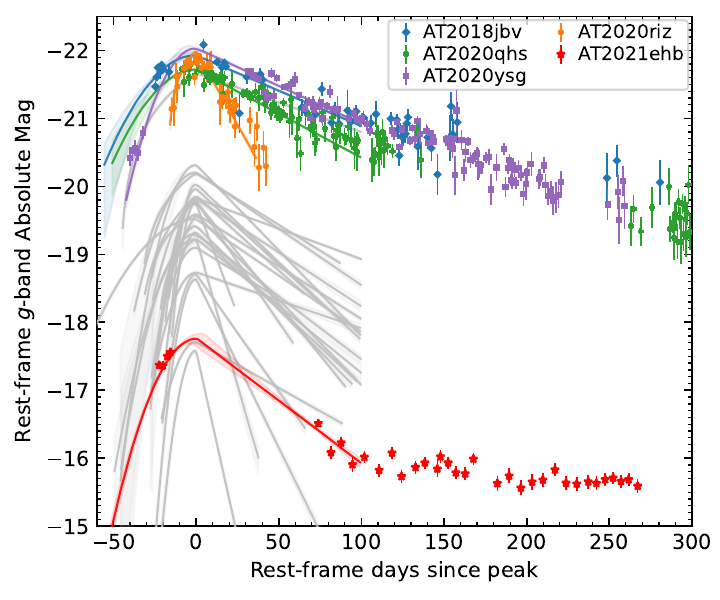}
    \caption{Rest-frame $g$-band light curve of \target compared with that of the 30 TDEs presented by \citet{Hammerstein2022}. The solid lines show the best-fit models (see text for details). 
    Data points and labeled names are only shown for TDEs with no discernible optical broad lines. \label{fig:lc_featureless}}
\end{figure}

Figure~\ref{fig:lc_featureless} compares the rest-frame $g$-band light curve of \target with 30 TDEs from phase-I of ZTF \citep{Hammerstein2022}. Solid lines are the results of fitting the multi-band light curves ($\delta t<100\,{\rm days}$) with a Gaussian rise + exponential decay model (see Section 5.1 of \citealt{vanVelzen2021}). We highlight the TDEs lacking line features by plotting the data as colored symbols. Here we have chosen an observing band with good temporal sampling, and converted the observations in this band into $\nu_0 = 6.3\times10^{14}\,{\rm Hz}$ by performing a color correction. 

Our study suggests that not all featureless TDEs are overluminous. In fact, the peak $g$-band magnitude ($M_{g, {\rm peak}}$) and peak $L_{\rm bb}$ of \target are faint compared with other optically selected TDEs (Figure~\ref{fig:bbfit}, Figure~\ref{fig:lc_featureless}). It is unclear whether $M_{g, {\rm peak}}$ of the TDE-featureless class forms a continuous or bimodal distribution between $-17$ and $-22$.
This question will be addressed in a forthcoming publication (Yao et al. in preparation).
A detailed analysis of Hubble Space Telescope UV spectroscopy  (Hammerstein et al. in preparation) will be essential to reveal if \target exhibits any spectral lines in the UV. 

\subsubsection{Origin of the NUV/optical Emission} \label{subsubsec:origin_UVopt}

Here we discuss possible origins of AT2021ehb’s NUV/optical emission: stream self-crossing shock, reprocessing, and thermal emission from disk accretion. 

In the self-crossing shock model, since the radius of the self-crossing shock is determined by the amount of general relativistic apsidal precession as given by the pericenter of the initial stellar orbit \citep{dai15_intersection_radius}, we expect the power of the self-crossing shock to track the fallback rate and decay with time as $\sim t^{-5/3}$. This is inconsistent with the flat light curve observed in AT2021ehb in the UV/optical bands (phase C--E), unless there is an additional mechanism that modulates the radiative efficiency of the self-crossing shock such that it roughly cancels the effects of the dropping shock dissipation power. Therefore, the energy dissipated by the stream-stream collision cannot be the primary source of emission during the plateau phase, although it may contribute to the early-time UV/optical emission.

In the reprocessing model, the nature of the reprocessing layer could originate from either a disk wind \citep{Strubbe2009, Miller2015_disk_wind, Dai2018, parkinson22_disk_wind, Thomsen2022_disk_wind} or an outflow from the self-crossing shock \citep{Jiang2016_self_crossing_shock, Lu2020}. The outflow scenario is favored with two reasons. 
First, a radiation pressure-driven disk wind originates from the innermost regions of the disk and the wind density is geometrically diluted as it propagates to a distance of the order $R_{\rm bb}\sim 10^{14}\,{\rm cm}$, whereas the outflow from the self-crossing shock is expected to be much denser near the self-crossing point and is hence more capable of reprocessing the hard emission from the disk \citep{Bonnerot2021_disk_formation}. 
Second, as the accretion flow goes from super-Eddington to sub-Eddington (\S\ref{subsubsec:thick-thin-disk}), one may expect the reduction in radiation pressure to reduce the strength of the wind outflows. 
The fact that the UV/optical luminosity only slightly decreases from phase D to phase E suggests that the reprocessing layer is not sensitive to the innermost accretion flow. 

Finally, if the UV/optical emission is powered by disk accretion, then the outer disk radius must be located at $\gtrsim 10R_{\rm T}\approx 100R_{\rm g}$. Recent simulations show that it is possible that a small fraction of the bound debris circularizes at $\sim 10R_{\rm T}$ \citep{Bonnerot2021_disk_formation}, but the accretion power at such large radii may be too low to produce the observed UV/optical emission, since the outermost regions of the disk is expected to be geometrically thin (due to efficient radiative cooling) with a very long viscous time. More detailed disk evolution modeling is needed to evaluate this possibility. 

To summarize, we infer that the early-time UV/optical light may be thermal radiation emitted at the photosphere of a stream-stream collision shock. The late-time UV/optical emission likely comes from reprocessing by the outflow launched from the self-crossing shock, although thermal emission from the outer regions of an accretion flow is not ruled out.

\subsection{Comparison to Other Accreting Black Holes} \label{subsec:disk_corona}

In stellar-mass black hole XRB outbursts, some objects are observed to transition between a soft disk-dominated state (SDS) and a hard Comptonized state (HCS). 
In the SDS of XRBs, the inner radius ($R_{\rm in, d}$) of an optically thick, geometrically thin disk stays around the ISCO of $R_{\rm ISCO}\sim {\rm a\ few}\times R_{\rm g}$. 
When the outbursts transition to the HCS, $\dot M_{\rm acc}$ decreases, $R_{\rm in, d}$ progressively moves outwards to $\sim {\rm few} \times  100 R_{\rm g}$, leaving a radiatively inefficient, advection-dominated accretion flow \citep{Yuan2005, Yuan2014}. At the same time, a region of hot corona is formed close to the BH \citep{Done2007}. 
For MBH accretors, Seyferts have been proposed as the high-$M_{\rm BH}$ analogs of XRBs in the SDS, whereas low-luminosity AGNs and low-ionization nuclear emission-line regions are considered similar to XRBs in the HCS \citep{Falcke2004}. 

\begin{figure}[htbp!]
    \centering
    \includegraphics[width=0.95\columnwidth]{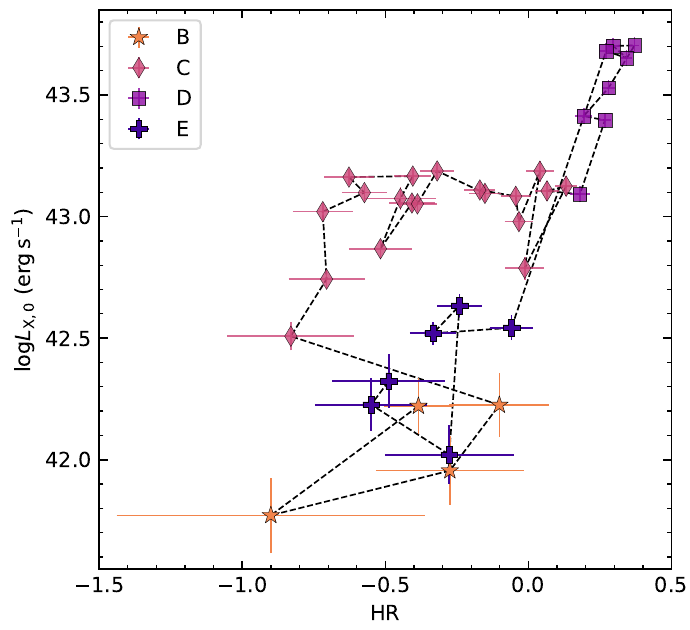}
    \caption{Galactic extinction-corrected 0.3--10\,keV X-ray luminosity ($L_{\rm X, 0}$) as a function of hardness ratio for \swift/XRT detections only. 
    \label{fig:Lx_HR}}
\end{figure}

In \S\ref{subsubsec:thick-thin-disk}, we propose that in phases B--D, the accretion flow of \target is in a radiation-trapped, super-Eddington regime (see, e.g., Fig.~2 of \citealt{Narayan2005}), which is different from the typical sub-Eddington X-ray states observed in AGNs and XRBs. 
Such a difference is further corroborated by two properties. 
First, on the hardness--intensity diagram (HID; see Figure~\ref{fig:Lx_HR}), the evolution of \target is neither similar to the ``turtlehead'' pattern observed in XRBs \citep{Fender2004, Munoz-Darias2013, Tetarenko2016}, nor similar to the ``brighter when softer'' trend observed in X-ray bright AGNs \citep{Auchettl2018}.
On the $L_{\rm bol}$--$\alpha_{\rm OX}$ diagram (Figure~\ref{fig:Lbol_alpha}), its evolution is also different from that observed in CLAGNs \citep{Ruan2019}. 
Second, in the canonical hard state (i.e. sub-Eddington accretion rates), there is a correlation among radio luminosity, X-ray luminosity, and $M_{\rm BH}$, which applies to both XRBs and hard-state AGN \citep{Merloni2003, Falcke2004}. 
A recent fit to this ``fundamental plane of black hole activity'' was provided by \citet{Gultekin2019}:
\begin{align}
    (1.09\pm0.10) R =& ({\rm log}M_7-1) - (0.55\pm0.22) \notag\\
    &- (-0.59_{-0.15}^{+0.16})X, \label{eq:fp}
\end{align}
where $R \equiv {\rm log}[L_{\rm 5\,{\rm GHz}}/(10^{38}\,{\rm erg\,s^{-1}})]$ and $X \equiv {\rm log}[L_{\rm 2-10\,{\rm keV}}/(10^{40}\,{\rm erg\,s^{-1}})]$.
In the hard state of \target, the 2--10\,keV luminosity in the \swift/XRT `s6' spectrum gives $X = 2.96\pm0.02$.
Using Eq.~(\ref{eq:fp}), the expected 5\,GHz radio luminosity on the fundamental plane is ${\rm log}[L_{\rm 5\,{\rm GHz}}/({\rm erg\,s^{-1}})]=R+38 = 38.21 \pm 0.59$.
The uncertainty of $R$ is calculated from the distribution of $10^5$ MC trials.
Using our radio limit at $\delta t=228$\,days (Table~\ref{tab:vla}) and assuming a flat radio spectrum of $f_\nu \propto \nu^0$, we find a 5\,GHz equivalent radio upper limit of ${\rm log}[L_{\rm 5\,{\rm GHz}}/({\rm erg\,s^{-1}})]<35.76$.
According to \citet{Gultekin2019}, the scatter on the fundamental plane correlation is a factor of $\sim7.6$ in the $L_{\rm 5\,{\rm GHz}}$ direction (or $\sim10$ in the $M_{\rm BH}$ direction), which is not enough to solve the discrepancy.
This again argues that \target is not in a canonical hard state where we would expect a radio jet to be launched.

The radio behavior of BH binaries at the Eddington accretion rates might be more relevant to \target (at least in phases B--D). 
A few XRBs get above the Eddington limit for brief periods, when they are seen to undergo a sequence of bright radio flares for a short period of time. 
Examples include the 2015 outburst of V404 Cygni \citep{Tetarenko2017, Tetarenko2019} and the ultraluminous X-ray source in M31 \citep{Middleton2013}.
In \target, the slow cadence of our radio follow-up observations does not allow us to rule out the existence of such radio flares, which last for hours to weeks, not months. 

Finally, we note that the evolution of the X-ray properties of \target are different from a few other TDEs. 
For example, \citet{Wevers2021} constructed the log$\lambda_{\rm Edd}$--$\alpha_{\rm OX}$ diagram for AT2018fyk, finding that the 2\,keV (corona) emission is stronger when $\lambda_{\rm Edd}$ is lower. 
Figure~\ref{fig:Lbol_alpha} shows that this is clearly not the case for \target. 
Separately, \citet{Hinkle2021} studied the evolution of AT2019azh on the canonical HID, showing that when the X-ray luminosity is higher, the X-ray spectrum is softer. 
Figure~\ref{fig:Lx_HR} shows that \target does not follow this trend either. 
It remains to be seen whether AT2021ehb is peculiar among the sample of optically selected TDEs with significant X-ray spectral evolution. To this end, constructing a systematically selected sample and analyzing the multi-wavelength data in a homogeneous fashion is the key.

\section{Conclusion} \label{sec:conclusion}
We have presented an extensive multi-wavelength study of the TDE AT2021ehb. Its peak 0.3--10\,keV flux of $\sim 1\,{\rm mCrab}$ is brighter than any other non-jetted TDE in the literature, allowing us to obtain a series of high-quality X-ray spectra, including the first hard X-ray spectrum of a non-jetted TDE up to 30\,keV. The detection of hard X-ray photons favors a highly aspherical geometry (\S\ref{subsec:hardXray}), and detailed modeling of the \nicer+\nustar spectrum shows evidence of relativistic disk reflection (\S\ref{subsubsec:joint2}). 

The emission from the self-crossing shock itself might contribute to the early-time UV/optical emission, while the post-peak (phase C--E) emission is either dominated by reprocessing of X-ray photons by the outflow launched from the shock, or by thermal emission in the outer regions of an accretion flow (\S\ref{subsubsec:origin_UVopt}). More detailed hydrodynamic and radiative transfer calculations \citep[e.g.,][]{Roth2016} are needed to test if such scenarios can reproduce the observed UV/optical plateau and the featureless optical spectra.

We observed a soft $\rightarrow$ hard $\rightarrow$ soft spectral transition in the X-ray. 
The initial soft-to-hard transition happened gradually over $\sim 170$\,days. 
A possible explanation is that magnetic fields grow with time due to differential rotation, resulting in the formation of a magnetically dominated hot corona (\S\ref{subsubsec:soft-to-hard}).
The bolometric luminosity of \target is the highest when the X-ray spectrum is the hardest --- a property that is different from XRBs, X-ray bright AGNs, and many other TDEs (\S\ref{subsec:disk_corona}). 
The latter hard-to-soft transition happened drastically within 3\,days at $\delta t \approx 272\,{\rm days}$, and might be due to thermal--viscous instability in the inner disk (\S\ref{subsubsec:thick-thin-disk}). Such an instability typically occurs when $L_{\rm bol} \sim 0.3 L_{\rm Edd}$ \citep{tchekhovskoy14_mad_jet}. This requires that most of the luminosity of \target is emitted in the EUV band that is not observed. 

Systems similar to \target are excellent targets for X-ray telescopes to study the real-time formation of the accretion disk and corona around MBHs. The detection of relativistic disk reflection features demonstrates the possibility of constraining  the spin of normally quiescent MBHs via reflection spectroscopy --- an opportunity enabled by combining modern time-domain surveys with systematic multi-wavelength follow-up observations.


\vspace{1cm}

\textit{Acknowledgements} -- 
We are grateful to the \nustar, \nicer, \swift, \xmm, and VLA teams for making this observing campaign possible. 
We thank the anonymous referee for constructive comments and suggestions.
We thank Julian Krolik for providing comments on an early version of this manuscript. We thank Erin Kara, Renee Ludlam, Guglielmo (Gullo) Mastroserio, and Riley Connors for helpful discussions on the \nustar and \nicer spectral fitting. 
We thank Murray Brightman and Hannah Earnshaw for discussions on disk reflection and super-Eddington accretion. 

Y.Y. acknowledges support from NASA under award No. 80NSSC22K0574. 
W.L. is supported by the Lyman Spitzer, Jr. Fellowship at Princeton University. 
This work was supported by the Australian government through the Australian Research Council's Discovery Projects funding scheme (DP200102471).
M.N. is supported by the European Research Council (ERC) under the European Union’s Horizon 2020 research and innovation program (grant agreement No.~948381) and by a Fellowship from the Alan Turing Institute.
E.C.K. acknowledges support from the G.R.E.A.T. research environment funded by {\em Vetenskapsr\aa det}, the Swedish Research Council, under project No. 2016-06012, and support from The Wenner-Gren Foundations.

This work has made use of data from the \nustar mission, a project led by Caltech, managed by NASA/JPL, and funded by NASA. 
This research has made use of the NuSTAR Data Analysis Software (NuSTARDAS), jointly developed by the ASI Science Data Center (ASDC, Italy) and Caltech (USA).

This work is based on observations obtained with the Samuel Oschin Telescope 48-inch and the 60-inch Telescope at the Palomar Observatory as part of the Zwicky Transient Facility project. ZTF is supported by the National Science Foundation under grant No. AST-2034437 and a collaboration including Caltech, IPAC, the Weizmann Institute of Science, the Oskar Klein Center at Stockholm University, the University of Maryland, Deutsches Elektronen-Synchrotron and Humboldt University, the TANGO Consortium of Taiwan, the University of Wisconsin at Milwaukee, Trinity College Dublin, Lawrence Livermore National Laboratories, IN2P3, University of Warwick, Ruhr University Bochum, and Northwestern University. Operations are conducted by COO, IPAC, and UW.
The ZTF forced-photometry service was funded under the Heising-Simons Foundation grant No. 12540303 (PI: Graham).

SED Machine is based upon work supported by the National Science Foundation under grant No. 1106171.

This work used observations with the eROSITA telescope on board the \srg observatory. 
The \srg observatory was built by Roskosmos with the participation of the Deutsches Zentrum f$\ddot{\rm u}$r Luft- und Raumfahrt (DLR). The \srg/eROSITA X-ray telescope was built by a consortium of German Institutes led by MPE, and supported by DLR.  The \srg spacecraft was designed, built, launched, and is operated by the Lavochkin Association and its subcontractors. The eROSITA data used in this work were processed using the eSASS software system developed by the German eROSITA consortium and proprietary data reduction and analysis software developed by the Russian eROSITA Consortium.

This work made use of data supplied by the UK Swift Science Data Centre at the University of Leicester.

The National Radio Astronomy Observatory is a facility of the National Science Foundation operated under cooperative agreement by Associated Universities, Inc.


\software{
\texttt{astropy} \citep{Astropy2013},
\texttt{CASA} \citep{McMullin2007}, 
\texttt{emcee} \citep{Foreman-Mackey2013},
\texttt{heasoft} \citep{HEASARC2014},
\texttt{LPipe} \citep{Perley2019lpipe}, 
\texttt{matplotlib} \citep{Hunter2007},
\texttt{Prospector} \citep{Johnson2021},
\texttt{pwkit} \citep{Williams2017}, 
\texttt{python-fsps} \citep{Foreman-Mackey2014},
\texttt{relxill} \citep{Garcia2014, Dauser2014},
\texttt{scipy} \citep{Virtanen2020},
\texttt{xspec} \citep{Arnaud1996}
}

\facilities{
XMM, 
NuSTAR,
Swift,
NICER,
SRG, 
PO:1.2m, 
PO:1.5m, 
Hale, 
Keck:I (LRIS), 
Keck: II (ESI),
LDT, 
VLA}

\clearpage
\appendix

\section{Supplementary Tables} \label{sec:tables}

\subsection{Photometry and Observing Logs} \label{subsec:obs_log}

UV and optical photometry are presented in Table~\ref{tab:phot}. 
\swift/XRT observations are summarized in Table~\ref{tab:xrt}.

\begin{deluxetable}{ccccc}[htbp!]
	\tablecaption{UV and optical photometry of \target.\label{tab:phot}}
	\tablehead{
		\colhead{MJD} &
		\colhead{Instrument} &
		\colhead{Filter} &
		\colhead{$f_\nu$ ($\mu$Jy)} &
		\colhead{$\sigma_{f_\nu}$ ($\mu$Jy)} 
	}
\startdata
59250.1643 & ZTF & $r$ & $-3.31$ & $12.48$ \\
59250.2031 & ZTF & $g$ & $2.07$ & $10.91$ \\
59299.0783 & UVOT & $uvw1$ & $567.68$ & $27.76$ \\
59299.0798 & UVOT & $U$ & $551.85$ & $41.60$ \\
59299.0808 & UVOT & $B$ & $487.52$ & $76.95$ \\
59299.0831 & UVOT & $uvw2$ & $587.34$ & $22.36$ \\
59299.0855 & UVOT & $V$ & $260.66$ & $146.91$ \\
59299.0875 & UVOT & $uvm2$ & $528.68$ & $19.49$ \\
\enddata
\tablecomments{$f_\nu$ is observed flux density before extinction correction. 
	(This table is available in its entirety in machine-readable form.)}
\end{deluxetable}

\begin{deluxetable*}{ccrrrrr}[htbp!]
\tabletypesize{\scriptsize}
    \tablecaption{Log of \swift/XRT observations of \target.\label{tab:xrt}}
	\tablehead{
	\colhead{obsID} 
    & \colhead{Start Date} 
	& \colhead{$\delta t$}
	& \colhead{Exp.}
	& \colhead{Net Count Rate}
	& \colhead{$f_{\rm X}$}
	& \colhead{$f_{\rm X, 0}$}
	\\
	\colhead{}
	& \colhead{}
	& \colhead{(days)}
	& \colhead{(s)} 
	& \colhead{($\rm count\,s^{-1}$)} 
	& \colhead{($10^{-13}\,{\rm erg\,s^{-1}\,cm^{-2}}$)}
	& \colhead{($10^{-13}\,{\rm erg\,s^{-1}\,cm^{-2}}$)}
    }
    \startdata
14217001 & 2021-03-26.0 & $-21.6$ & 2669 & $<0.0019$ & $<0.66$ & $<1.25$ \\
14217003 & 2021-03-28.2 & $-19.4$ & 1475 & $<0.0027$ & $<0.96$ & $<1.82$ \\
14217004 & 2021-03-31.0 & $-16.7$ & 1683 & $<0.0024$ & $<0.84$ & $<1.59$ \\
14217005 & 2021-04-02.0 & $-14.7$ & 1336 & $<0.0030$ & $<1.06$ & $<2.01$ \\
\hline
14217006 & 2021-07-01.2 & $+73.9$ & 4078 & $0.0339 \pm 0.0029$ & $12.01 \pm 3.21$ & $22.73 \pm 6.08$ \\
14217007 & 2021-07-09.8 & $+82.3$ & 1366 & $0.0120 \pm 0.0030$ & $4.27 \pm 1.52$ & $8.08 \pm 2.88$ \\
14217008 & 2021-07-16.1 & $+88.5$ & 1348 & $0.0184 \pm 0.0037$ & $6.52 \pm 2.11$ & $12.34 \pm 4.00$ \\
14217009 & 2021-07-23.1 & $+95.4$ & 1141 & $0.0343 \pm 0.0056$ & $12.13 \pm 3.65$ & $22.96 \pm 6.90$ \\
\hline
14217010 & 2021-07-30.1 & $+102.3$ & 1366 & $0.0502 \pm 0.0061$ & $16.57 \pm 2.23$ & $44.04 \pm 5.92$ \\
14217011 & 2021-08-08.1 & $+111.1$ & 1925 & $0.0863 \pm 0.0067$ & $28.44 \pm 2.76$ & $75.62 \pm 7.34$ \\
14217012 & 2021-08-15.9 & $+118.8$ & 1653 & $0.1635 \pm 0.0100$ & $53.90 \pm 4.54$ & $143.32 \pm 12.08$ \\
14217013 & 2021-08-22.1 & $+124.8$ & 2065 & $0.1958 \pm 0.0098$ & $64.56 \pm 4.94$ & $171.64 \pm 13.13$ \\
14217014 & 2021-08-30.9 & $+133.5$ & 1583 & $0.2268 \pm 0.0120$ & $74.78 \pm 5.87$ & $198.82 \pm 15.61$ \\
\hline
14217015 & 2021-09-05.5 & $+139.0$ & 1830 & $0.2548 \pm 0.0119$ & $94.07 \pm 7.27$ & $200.69 \pm 15.51$ \\
14217016 & 2021-09-12.8 & $+146.2$ & 641 & $0.2061 \pm 0.0180$ & $76.09 \pm 8.13$ & $162.34 \pm 17.35$ \\
14217017 & 2021-09-15.0 & $+148.4$ & 1503 & $0.1281 \pm 0.0093$ & $47.29 \pm 4.51$ & $100.90 \pm 9.63$ \\
14217018 & 2021-09-19.7 & $+153.0$ & 1580 & $0.1974 \pm 0.0112$ & $72.90 \pm 6.12$ & $155.52 \pm 13.05$ \\
14217019 & 2021-09-24.2 & $+157.4$ & 2045 & $0.1959 \pm 0.0098$ & $72.33 \pm 5.76$ & $154.31 \pm 12.28$ \\
14217020 & 2021-09-30.4 & $+163.5$ & 1867 & $0.2675 \pm 0.0120$ & $98.74 \pm 7.54$ & $210.67 \pm 16.09$ \\
\hline
14217021 & 2021-10-05.5 & $+168.5$ & 1595 & $0.2775 \pm 0.0132$ & $104.81 \pm 9.01$ & $170.40 \pm 14.65$ \\
14217022 & 2021-10-20.2 & $+182.9$ & 1618 & $0.2865 \pm 0.0134$ & $108.22 \pm 9.24$ & $175.94 \pm 15.02$ \\
14217023 & 2021-10-27.4 & $+190.0$ & 1480 & $0.2698 \pm 0.0136$ & $101.93 \pm 8.91$ & $165.71 \pm 14.48$ \\
14217024 & 2021-11-03.5 & $+197.0$ & 2010 & $0.2124 \pm 0.0103$ & $80.23 \pm 6.94$ & $130.44 \pm 11.29$ \\
\hline
14217025 & 2021-11-10.7 & $+204.0$ & 1286 & $0.3132 \pm 0.0157$ & $149.64 \pm 15.02$ & $210.15 \pm 21.10$ \\
14217026 & 2021-11-17.2 & $+210.4$ & 1813 & $0.1251 \pm 0.0084$ & $59.75 \pm 6.57$ & $83.92 \pm 9.23$ \\
14217027 & 2021-11-24.7 & $+217.7$ & 1957 & $0.2718 \pm 0.0119$ & $129.86 \pm 12.64$ & $182.38 \pm 17.75$ \\
14217028 & 2021-12-01.5 & $+224.4$ & 1967 & $0.2600 \pm 0.0116$ & $124.20 \pm 12.14$ & $174.42 \pm 17.05$ \\
\hline
14217029 & 2021-12-08.1 & $+230.9$ & 2317 & $0.2596 \pm 0.0107$ & $126.84 \pm 9.59$ & $168.93 \pm 12.77$ \\
14217030 & 2021-12-15.2 & $+237.9$ & 2010 & $0.5234 \pm 0.0162$ & $255.79 \pm 18.06$ & $340.66 \pm 24.05$ \\
14217031 & 2021-12-20.3 & $+242.9$ & 1293 & $0.5445 \pm 0.0206$ & $266.11 \pm 19.65$ & $354.40 \pm 26.17$ \\
14217032 & 2021-12-25.6 & $+248.2$ & 1395 & $0.7108 \pm 0.0227$ & $347.35 \pm 24.66$ & $462.59 \pm 32.84$ \\
\hline
14217033 & 2021-12-30.5 & $+253.0$ & 1371 & $0.9721 \pm 0.0268$ & $551.93 \pm 41.79$ & $691.67 \pm 52.37$ \\
14217034 & 2022-01-04.5 & $+257.8$ & 1410 & $0.9675 \pm 0.0263$ & $549.33 \pm 41.53$ & $688.41 \pm 52.05$ \\
14217035 & 2022-01-09.2 & $+262.4$ & 1361 & $0.8629 \pm 0.0253$ & $489.92 \pm 37.43$ & $613.96 \pm 46.91$ \\
14217036 & 2022-01-14.7 & $+267.9$ & 1423 & $0.9218 \pm 0.0256$ & $523.38 \pm 39.68$ & $655.88 \pm 49.72$ \\
\hline
14217041 & 2022-02-23.1 & $+306.6$ & 2594 & $0.0745 \pm 0.0054$ & $26.05 \pm 3.06$ & $47.73 \pm 5.60$ \\
14217042 & 2022-03-02.2 & $+313.5$ & 3888 & $0.0706 \pm 0.0043$ & $24.72 \pm 2.73$ & $45.29 \pm 5.01$ \\
14217043 & 2022-03-09.7 & $+320.9$ & 2766 & $0.0918 \pm 0.0058$ & $32.11 \pm 3.59$ & $58.83 \pm 6.58$ \\
\hline
14217044 & 2022-03-16.1 & $+327.2$ & 2956 & $0.0122 \pm 0.0022$ & $5.03 \pm 1.40$ & $14.35 \pm 3.98$ \\
14217045 & 2022-03-23.0 & $+334.0$ & 3263 & $0.0197 \pm 0.0025$ & $8.08 \pm 2.01$ & $23.04 \pm 5.73$ \\
14217046 & 2022-03-30.5 & $+341.3$ & 2354 & $0.0246 \pm 0.0033$ & $10.11 \pm 2.55$ & $28.81 \pm 7.26$ \\
    \enddata
\tablecomments{All measurements are given in 0.3--10\,keV. $f_{\rm X}$ and $f_{\rm X, 0}$ are converted using the scaling factors derived in Table~\ref{tab:xrtfit}. }
\end{deluxetable*}

\subsection{Model Fits} \label{subsec:modelfit}
Table~\ref{tab:bbpars} presents the photospheric parameters derived from fitting a blackbody function to the UV/optical data. Note that the uncertainty of $T_{\rm bb}$ is only computed for ``good'' epochs (see details in \S\ref{subsec:uvopt_bbfit}).

\begin{deluxetable*}{rccc}[htbp!]
\tabletypesize{\scriptsize}
    \tablecaption{UV/optical blackbody parameters.\label{tab:bbpars}}
    \tablehead{
    \colhead{$\delta t$ (days)}
    & \colhead{$L_{\rm bb}$ ($10^{43}\,{\rm erg\,s^{-1}}$)}
    & \colhead{$R_{\rm bb}$ ($10^{14}\,{\rm cm}$)}
    & \colhead{$T_{\rm bb}$ ($10^3\,{\rm K}$)}
    }
    \startdata
\ad{$-50.0$} & \ad{$0.39 \pm 0.15$} & \ad{$0.91 \pm 0.03$} & \ad{$2.85$} \\
\ad{$-45.0$} & \ad{$0.67 \pm 0.25$} & \ad{$1.19 \pm 0.01$} & \ad{$2.85$} \\
\ad{$-40.0$} & \ad{$1.01 \pm 0.38$} & \ad{$1.46 \pm 0.05$} & \ad{$2.85$} \\
\ad{$-30.0$} & \ad{$2.10 \pm 0.78$} & \ad{$2.11 \pm 0.04$} & \ad{$2.85$} \\
\ad{$-25.0$} & \ad{$2.62 \pm 0.96$} & \ad{$2.36 \pm 0.03$} & \ad{$2.85$} \\
\ad{$-21.5$} & \ad{$2.77 \pm 1.19$} & \ad{$2.42 \pm 0.27$} & \ad{$2.85 \pm 0.26$} \\
\ad{$-19.2$} & \ad{$2.92 \pm 1.30$} & \ad{$2.51 \pm 0.29$} & \ad{$2.84 \pm 0.27$} \\
\ad{$-16.7$} & \ad{$3.00 \pm 0.89$} & \ad{$2.70 \pm 0.20$} & \ad{$2.75 \pm 0.18$} \\
\ad{$-14.7$} & \ad{$3.13 \pm 0.65$} & \ad{$2.80 \pm 0.14$} & \ad{$2.73 \pm 0.13$} \\
\ad{$-10.0$} & \ad{$3.31 \pm 0.62$} & \ad{$2.89 \pm 0.02$} & \ad{$2.73$} \\
\ad{$70.0$} & \ad{$1.33 \pm 0.29$} & \ad{$1.91 \pm 0.02$} & \ad{$2.67$} \\
\ad{$74.7$} & \ad{$1.12 \pm 0.29$} & \ad{$1.75 \pm 0.12$} & \ad{$2.67 \pm 0.15$} \\
\ad{$82.4$} & \ad{$1.06 \pm 0.86$} & \ad{$1.45 \pm 0.31$} & \ad{$2.90 \pm 0.50$} \\
\ad{$88.6$} & \ad{$1.27 \pm 1.46$} & \ad{$1.18 \pm 0.34$} & \ad{$3.36 \pm 0.84$} \\
\ad{$95.4$} & \ad{$0.79 \pm 0.67$} & \ad{$1.42 \pm 0.33$} & \ad{$2.72 \pm 0.48$} \\
\ad{$102.3$} & \ad{$0.55 \pm 0.18$} & \ad{$1.79 \pm 0.17$} & \ad{$2.22 \pm 0.15$} \\
\ad{$111.5$} & \ad{$0.52 \pm 0.12$} & \ad{$1.70 \pm 0.10$} & \ad{$2.24 \pm 0.11$} \\
\ad{$119.3$} & \ad{$0.60 \pm 0.10$} & \ad{$1.56 \pm 0.07$} & \ad{$2.43 \pm 0.09$} \\
\ad{$125.1$} & \ad{$0.54 \pm 0.20$} & \ad{$1.67 \pm 0.18$} & \ad{$2.28 \pm 0.18$} \\
\ad{$133.6$} & \ad{$0.50 \pm 0.18$} & \ad{$1.82 \pm 0.19$} & \ad{$2.14 \pm 0.16$} \\
\ad{$139.2$} & \ad{$0.52 \pm 0.13$} & \ad{$1.80 \pm 0.12$} & \ad{$2.18 \pm 0.11$} \\
\ad{$147.0$} & \ad{$0.57 \pm 0.15$} & \ad{$1.65 \pm 0.11$} & \ad{$2.32 \pm 0.13$} \\
\ad{$153.3$} & \ad{$0.60 \pm 0.25$} & \ad{$1.51 \pm 0.18$} & \ad{$2.47 \pm 0.22$} \\
\ad{$157.9$} & \ad{$0.57 \pm 0.30$} & \ad{$1.47 \pm 0.21$} & \ad{$2.47 \pm 0.27$} \\
\ad{$163.6$} & \ad{$0.54 \pm 0.19$} & \ad{$1.50 \pm 0.15$} & \ad{$2.41 \pm 0.18$} \\
\ad{$168.6$} & \ad{$0.63 \pm 0.25$} & \ad{$1.31 \pm 0.14$} & \ad{$2.68 \pm 0.23$} \\
\ad{$178.0$} & \ad{$0.42 \pm 0.18$} & \ad{$1.65 \pm 0.18$} & \ad{$2.16 \pm 0.20$} \\
\ad{$183.1$} & \ad{$0.44 \pm 0.25$} & \ad{$1.51 \pm 0.25$} & \ad{$2.28 \pm 0.26$} \\
\ad{$190.2$} & \ad{$0.43 \pm 0.22$} & \ad{$1.55 \pm 0.23$} & \ad{$2.24 \pm 0.23$} \\
\ad{$197.3$} & \ad{$0.37 \pm 0.13$} & \ad{$1.76 \pm 0.17$} & \ad{$2.02 \pm 0.14$} \\
\ad{$204.1$} & \ad{$0.43 \pm 0.19$} & \ad{$1.53 \pm 0.18$} & \ad{$2.25 \pm 0.20$} \\
\ad{$210.5$} & \ad{$0.44 \pm 0.25$} & \ad{$1.59 \pm 0.26$} & \ad{$2.22 \pm 0.26$} \\
\ad{$217.8$} & \ad{$0.44 \pm 0.20$} & \ad{$1.56 \pm 0.05$} & \ad{$2.24$} \\
\ad{$224.5$} & \ad{$0.42 \pm 0.18$} & \ad{$1.51 \pm 0.07$} & \ad{$2.26$} \\
\ad{$231.0$} & \ad{$0.40 \pm 0.17$} & \ad{$1.44 \pm 0.05$} & \ad{$2.27$} \\
\ad{$238.0$} & \ad{$0.42 \pm 0.17$} & \ad{$1.47 \pm 0.04$} & \ad{$2.29$} \\
\ad{$242.9$} & \ad{$0.42 \pm 0.16$} & \ad{$1.45 \pm 0.05$} & \ad{$2.30$} \\
\ad{$248.2$} & \ad{$0.42 \pm 0.16$} & \ad{$1.43 \pm 0.05$} & \ad{$2.32$} \\
\ad{$253.0$} & \ad{$0.43 \pm 0.16$} & \ad{$1.43 \pm 0.04$} & \ad{$2.33$} \\
\ad{$257.8$} & \ad{$0.43 \pm 0.15$} & \ad{$1.42 \pm 0.03$} & \ad{$2.34$} \\
\ad{$262.4$} & \ad{$0.42 \pm 0.14$} & \ad{$1.39 \pm 0.03$} & \ad{$2.35$} \\
\ad{$267.9$} & \ad{$0.41 \pm 0.13$} & \ad{$1.35 \pm 0.03$} & \ad{$2.37$} \\
\ad{$276.0$} & \ad{$0.39 \pm 0.12$} & \ad{$1.29 \pm 0.04$} & \ad{$2.39$} \\
\ad{$296.0$} & \ad{$0.42 \pm 0.11$} & \ad{$1.29 \pm 0.03$} & \ad{$2.44$} \\
\ad{$302.0$} & \ad{$0.38 \pm 0.11$} & \ad{$1.22 \pm 0.08$} & \ad{$2.45 \pm 0.15$} \\
\ad{$306.7$} & \ad{$0.39 \pm 0.10$} & \ad{$1.22 \pm 0.02$} & \ad{$2.45$} \\
\ad{$313.7$} & \ad{$0.39 \pm 0.10$} & \ad{$1.23 \pm 0.04$} & \ad{$2.45$} \\
\ad{$320.9$} & \ad{$0.32 \pm 0.08$} & \ad{$1.11 \pm 0.04$} & \ad{$2.45$} \\
\ad{$327.5$} & \ad{$0.26 \pm 0.07$} & \ad{$1.01 \pm 0.03$} & \ad{$2.45$} \\
\ad{$334.3$} & \ad{$0.24 \pm 0.06$} & \ad{$0.97 \pm 0.03$} & \ad{$2.45$} \\
\ad{$341.4$} & \ad{$0.25 \pm 0.06$} & \ad{$0.98 \pm 0.03$} & \ad{$2.45$} \\
    \enddata
\end{deluxetable*}

The XRT spectral parameters are presented in Table~\ref{tab:xrtfit}.

\begin{deluxetable*}{ccccccc}[htbp!]
\tabletypesize{\scriptsize}
    \tablecaption{X-ray Fluxes from Modeling of XRT spectra.\label{tab:xrtfit}}
	\tablehead{
	\colhead{Observation}
	& \colhead{Net 0.3--10\,keV Rate}
	& \colhead{$f_\nu({\rm 0.5\,keV})$}
	& \colhead{$f_\nu({\rm 2\,keV})$}
	& \colhead{\ad{$f_{\rm X}$ (0.3--10\,keV)}}
	& \colhead{\ad{$f_{\rm X, 0}$ (0.3--10\,keV)}}
	& \colhead{\ad{$f_{\rm X, 0}$ (0.5--10\,keV)}}\\
	\colhead{}
	& \colhead{($\rm count\,s^{-1}$)}
	& \colhead{(\ad{$\mu$Jy})}
	& \colhead{(\ad{$\mu$Jy})}
	& \colhead{($10^{-13}\,{\rm erg\,s^{-1}\,cm^{-2}}$)}
	& \colhead{($10^{-13}\,{\rm erg\,s^{-1}\,cm^{-2}}$)}
	& \colhead{($10^{-13}\,{\rm erg\,s^{-1}\,cm^{-2}}$)}
	}
	\startdata
s1 & $0.0276 \pm 0.0019$ & \ad{$1.133_{-0.185}^{+0.103}$} & \ad{$0.030_{-0.023}^{+0.002}$} & $9.76_{-1.08}^{+1.39}$ & $18.47_{-2.05}^{+2.63}$ & $11.34_{-2.72}^{+0.37}$ \\
s2 & $0.1476 \pm 0.0041$ & \ad{$9.639_{-0.595}^{+0.178}$} & \ad{$0.106_{-0.012}^{+0.008}$} & $48.67_{-1.03}^{+1.79}$ & $129.40_{-2.74}^{+4.76}$ & $48.96_{-2.53}^{+1.34}$ \\
s3 & $0.2116 \pm 0.0047$ & \ad{$10.930_{-0.694}^{+0.287}$} & \ad{$0.312_{-0.018}^{+0.017}$} & $78.12_{-1.95}^{+2.87}$ & $166.66_{-4.15}^{+6.13}$ & $86.04_{-3.95}^{+1.75}$ \\
s4 & $0.2584 \pm 0.0062$ & \ad{$7.565_{-0.683}^{+0.224}$} & \ad{$0.544_{-0.035}^{+0.026}$} & $97.63_{-3.31}^{+3.67}$ & $158.72_{-5.37}^{+5.97}$ & $109.87_{-5.73}^{+2.56}$ \\
s5 & $0.2382 \pm 0.0058$ & \ad{$5.485_{-0.583}^{+0.211}$} & \ad{$0.678_{-0.054}^{+0.029}$} & $113.80_{-4.00}^{+5.90}$ & $159.82_{-5.62}^{+8.29}$ & $127.47_{-7.81}^{+3.50}$ \\
s6 & $0.4776 \pm 0.0083$ & \ad{$8.675_{-1.257}^{+0.247}$} & \ad{$1.702_{-0.063}^{+0.057}$} & $233.38_{-4.48}^{+8.72}$ & $310.81_{-5.97}^{+11.61}$ & $256.24_{-11.39}^{+7.73}$ \\
s7 & $0.9314 \pm 0.0129$ & \ad{$14.647_{-1.061}^{+0.724}$} & \ad{$3.575_{-0.164}^{+0.082}$} & $528.81_{-13.21}^{+20.29}$ & $662.69_{-16.55}^{+25.43}$ & $579.24_{-27.10}^{+11.66}$ \\
s8 & $0.0780 \pm 0.0029$ & \ad{$2.843_{-0.351}^{+0.121}$} & \ad{$0.132_{-0.014}^{+0.010}$} & $27.30_{-0.96}^{+1.56}$ & $50.03_{-1.77}^{+2.85}$ & $30.99_{-2.74}^{+1.06}$ \\
s9 & $0.0185 \pm 0.0015$ & \ad{$1.218_{-0.741}^{+0.006}$} & \ad{$0.027_{-0.005}^{+0.006}$} & $7.59_{-0.52}^{+1.11}$ & $21.62_{-1.48}^{+3.15}$ & $6.78_{-1.38}^{+0.48}$ \\
	\enddata
\end{deluxetable*}

\section{Optical Spectroscopy Instrumental/Observational Information} \label{sec:optspec_details}
A log of optical spectroscopic observation is given in Table~\ref{tab:spec}. 

\begin{deluxetable*}{crccccr}[htbp!]
\tabletypesize{\scriptsize}
	\tablecaption{Log of \target optical spectroscopy. \label{tab:spec}}
	\tablehead{
		\colhead{Start Date}  
		& \colhead{$\delta t$ (days)}
		& \colhead{Telescope}
		& \colhead{Instrument}
		& \colhead{Wavelength range (\AA)}
		& \colhead{Slit width ($^{\prime\prime}$)}
		& \colhead{Exp. (s)} 
	}
	\startdata
2021-03-25.1 & $-22$ & P60 & SEDM & 3770--9223 & --- & 2160 \\
2021-03-27.1 & $-20$ & P60 & SEDM & 3770--9223 & --- & 2160 \\
2021-07-06.6 & $+79$ & Keck-I & LRIS & 3200--10250 & 1.0 & 300 \\
2021-08-01.4 & $+104$ & P200 & DBSP & 3410--5550, 5750--9995 & 1.5 & 900 \\
2021-08-13.6 & $+116$ & Keck-I & LRIS & 3200--10250 & 1.0 & 300 \\
2021-09-07.6 & $+141$ & Keck-I & LRIS & 3200--10250 & 1.0 & 300 \\
2021-09-17.4 & $+150$ & P60 & SEDM & 3770--9223 & --- & 2700 \\
2021-10-27.5 & $+190$ & LDT & DeVeny & 3586–8034 & 1.5 & 2400 \\
2021-11-13.3 & $+206$ & P60 & SEDM & 3770--9223 & --- & 2700 \\
2021-12-03.3 & $+226$ & P60 & SEDM & 3770--9223 & --- & 2700 \\
2021-12-28.4 & $+250$ & Keck-II & ESI & 4000--10250 & 0.75 & 300 \\
2022-01-05.2 & $+258$ & P60 & SEDM & 3770--9223 & --- & 2700 \\
2022-01-12.2 & $+265$ & P200 & DBSP & 3410--5550, 5750--9995 & 2.0 & 600 \\
2022-01-20.3 & $+273$ & P60 & SEDM & 3770--9223 & --- & 2700 \\
2022-01-27.3 & $+280$ & P60 & SEDM & 3770--9223 & --- & 2700 \\
2022-02-06.3 & $+290$ & Keck-I & LRIS & 3200--10250 & 1.0 & 300 \\
2022-03-27.1 & $+338$ & P200 & DBSP & 3410--5550, 5750--9995 & 1.5 & 1200 \\
\enddata 
\tablecomments{All spectra will be made available on the TNS page of this source (\url{https://www.wis-tns.org/object/2021ehb}) at the time of manuscript submission. }
\end{deluxetable*}

For LRIS observations, we used the 560 dichroic, the 400/3400 grism on the blue side, the 400/8500 grating on the red side, and the $1^{\prime\prime}$ slit width,
which gives $\sigma_{\rm inst}\approx173\,{\rm km\,s^{-1}}$ on the blue side and $\sigma_{\rm inst}\approx126\,{\rm km\,s^{-1}}$ on the red side. The LRIS spectra were reduced and extracted using \texttt{Lpipe} \citep{Perley2019lpipe}.

For DBSP observations, we used the D-55 dichroic filter, the 600/4000 grating on the blue side, the 316/7500 grating on the red side.
With a slit width of $1.5^{\prime\prime}$ ($2.0^{\prime\prime}$), this gives $\sigma_{\rm inst}\approx106\,{\rm km\,s^{-1}}$ ($\sigma_{\rm inst}\approx141\,{\rm km\,s^{-1}}$) on the blue side and $\sigma_{\rm inst}\approx 143\,{\rm km\,s^{-1}}$ ($\sigma_{\rm inst}\approx 190\,{\rm km\,s^{-1}}$) on the red side.
The DBSP spectra were reduced using the \texttt{dbsp\_drp} pipeline \citep{Roberson2022}, which is based on \texttt{PypeIt} \citep{pypeit:joss_pub}. 

The ESI observation was performed in the Echellette mode with a 0.75$^{\prime\prime}$ wide slit, which gives a resolving power of $R=5350$ (i.e., $\sigma_{\rm inst} = 24\,{\rm km\,s^{-1}}$).
The ESI spectrum was reduced using the \texttt{MAKEE} 
pipeline following standard procedures. Flux calibration was not performed. 
We normalized the spectra by fitting third-order cubic splines to the continuum, with prominent emission and absorption lines masked.  

Observations with DeVeny were performed with the 300/4000 grating, with a grating tilt angle of 23.13$\degree$ to yield a central wavelength of 5800 \AA, the clear rear filter, and a slit width of 1.5$\arcsec$. This gives $\sigma_{\rm inst} \approx 169\,{\rm km\,s^{-1}}$. DeVeny spectra were reduced with \texttt{PyRAF}, including bias correction and flat-fielding.

\bibliography{main}{}
\bibliographystyle{aasjournal}

\end{document}